\documentclass[aps,prl,twocolumn,showpacs,superscriptaddress,groupedaddress]{revtex4}  
\usepackage{graphicx}  
\usepackage{dcolumn}   
\usepackage{bm}        
\usepackage{amssymb}   
\usepackage{mathrsfs}
\usepackage{amssymb}
\usepackage{amsmath}

\begin{document}
\title{Phase transitions and spin excitations of spin-1 bosons in optical lattice}
\author{Min-Jie Zhu}
\affiliation{Shanghai Branch, National Laboratory for Physical Sciences at Microscale and
Department of Modern Physics, University of Science and Technology of China,
Hefei, Anhui 230026, China}
\affiliation{CAS Center for Excellence and Synergetic Innovation Center in Quantum
Information and Quantum Physics, University of Science and Technology of
China, Shanghai 201315, China}
\affiliation{CAS-Alibaba Quantum Computing Laboratory, Shanghai 201315, China}

\author{Bo Zhao}
\affiliation{Shanghai Branch, National Laboratory for Physical Sciences at Microscale and
Department of Modern Physics, University of Science and Technology of China,
Hefei, Anhui 230026, China}
\affiliation{CAS Center for Excellence and Synergetic Innovation Center in Quantum
Information and Quantum Physics, University of Science and Technology of
China, Shanghai 201315, China}
\affiliation{CAS-Alibaba Quantum Computing Laboratory, Shanghai 201315, China}

\begin{abstract}
We investigate ground state properties of spin-1 bosonic system trapped in optical lattice with extended standard basis operator (SBO) method. For both ferromagnetic ($U_2<0$) and antiferromagnetic ($U_2>0$) systems, we analytically figure out the symmetry properties in Mott-insulator and superfluid phases, which would provide a deeper insight into the MI-SF phase transition process. Then by applying self-consistent approach to the method, we include the effect of quantum and thermal fluctuations and derive the MI-SF transition phase diagram, which is in quantitative agreement with recent Monte-Carlo simulation at zero temperature, and at finite temperature, we find the underestimation of finite-temperature-effect in the mean-field approximation method. If we further consider the spin excitations in the insulating states of spin-1 system in external field, distinct spin phases are expected. Therefore, in the Mott lobes with $n=1$ and $n=2$ atoms per site, we give analytical and numerical boundaries of the singlet, nematic, partially magnetic and ferromagnetic phases in the magnetic phase diagrams.
\end{abstract}

\maketitle


\section{Introduction}

Ultracold atoms trapped in optical lattices provide clean and highly controllable platform for the study of strong-correlated systems~\cite{RevModPhys.80.885}. Besides the well-known Mott insulator and superfluid phases~\cite{PhysRevLett.81.3108,greiner2002quantum}, recent experiments are approaching the regime where magnetic properties could be revealed~\cite{PhysRevLett.104.180401,PhysRevLett.106.195301,PhysRevA.93.063607}, such as the expected Ne\'{e}l antiferromagnet in spin-1/2 fermions. The interplay of magnetism and superfluidity in such systems can be realized with spinor gases trapped in optical lattice. The presence of extra spin degrees of freedom induces exceedingly rich physics in spinor system and allows the studies of multi-band condensed matter Hamiltonians and symmetry and topology in quantum-ordered materials~\cite{RevModPhys.85.1191}.

Most work in spinor systems, both theoretically and experimentally~\cite{PhysRevLett.81.4531,PhysRevLett.81.243,PhysRevLett.95.190405,1367-2630-8-8-152,PhysRevLett.81.742,Ohmi1998Bose,PhysRevLett.110.130405,PhysRevB.69.094410,PhysRevLett.88.163001,PhysRevA.68.063602,PhysRevA.70.043628,PhysRevLett.94.110403,PhysRevB.77.014503,PhysRevLett.97.180411,PhysRevB.76.014502, PhysRevA.77.033628,PhysRevA.93.063607,PhysRevA.91.043620,So2016Strongly}, was focused on atoms with effective spin $F=1$, which is also exclusively considered in this paper. In theory, the spin-1 atoms trapped in optical lattice can be described by extended Bose-Hubbard model with spin-dependent terms~\cite{PhysRevA.68.063602,PhysRevLett.88.163001,RevModPhys.85.1191}, in which both the MI phases and MI-SF phase transitions have been intensively studied. For ferromagnetic interacting system ($U_2<0$), the whole phase diagram, including Mott insulator and superfluid phases, is predicted to be ferromagnetic~\cite{PhysRevLett.110.130405,PhysRevB.88.104509}, and all the Mott lobes shrink, eventually disappear, as the increase of spin-dependent interaction strength $|U_2|$~\cite{PhysRevLett.102.140402}. While for antiferromagnetic interacting system ($U_2>0$), the superfluid phase in the MI-SF transition is expected to be polarized with zero global magnetization in the whole phase diagram~\cite{PhysRevLett.88.163001,PhysRevLett.110.130405,PhysRevB.69.094410,PhysRevA.70.043628,PhysRevB.88.104509}, and the Mott lobes with even filling numbers grow at the expense of odd ones~\cite{PhysRevA.70.043628,PhysRevB.88.104509}. Then for small positive $U_2$, mean-field studies have pointed out that the MI-SF transition from even filling Mott lobes is first-order phase transition process (second-order for all other cases)~\cite{PhysRevB.69.094410,PhysRevLett.94.110403}, which has also been confirmed in experiment~\cite{PhysRevA.93.063607} and one-dimensional quantum Monte-Carlo simulation~\cite{PhysRevLett.102.140402}. Moreover, spin phases including singlet, nematic and dimerized phases are predicted in the Mott lobes for antiferromagnetic case~\cite{PhysRevLett.88.163001,PhysRevA.68.063602,PhysRevLett.93.120405}. If the filling number is odd, the Mott lobes are expected to form nematic state, which can be changed to dimerized state in one-dimension system~\cite{PhysRevA.68.063602,PhysRevLett.95.240404,PhysRevA.74.035601}. If the filling number is even, the Mott lobes are expected to have spin singlets in each site with a first-order transition to nematic phase when $zU_2/U_0<0.1$ ($z=2\text{dim}$)~\cite{PhysRevA.68.063602}.

If external magnetic field is applied to the system, numerous new physics can be found. For example, the transitions for bosons in Mott lobes with different filling numbers may occur into different superfluid phases with spins aligned along or opposite to the field direction~\cite{PhysRevA.68.043612}. In the insulating states, 3D mean-field analysis~\cite{PhysRevA.80.053615} predicted nematic to ferromagnetic (or partially magnetic) transitions, then in Ref.~\cite{PhysRevLett.106.105302}, the whole magnetic phase diagram is shown in the system at unit filling with quadratic Zeeman effect. In spite of these findings, a general analytical theory of the field-induced quantum phases in spin-1 bosonic system is still missing.

In this paper, we would like to propose extended standard basis operator (SBO) method to solve the spin-1 bosonic system trapped in optical lattice. This method, which is applied to the motion equation for Green's functions~\cite{PhysRevB.5.1106,PhysRevA.92.013602}, has the capacity to analytically solve most of the phase transition problems in spin-1 system on one hand, on the other hand, it is possible to handle quantum fluctuations in the system by taking self-consistent approach, showing that the method goes significantly beyond mean-field level. Therefore, we apply it to study the symmetry properties (polar or ferromagnetic) of superfluid and Mott-insulator phases in both ferromagnetic ($U_2<0$) and antiferromagnetic ($U_2>0$) systems. The MI-SF transition phase diagrams are given afterwards, and we find the self-consistent SBO phase diagrams at zero temperature are in quantitative agreement with quantum Monte-Carlo simulation results. Then in the Mott lobes of the system in external magnetic field, of which the effective Hamiltonian is derived with perturbation theory, we consider the spin excitations and calculate the excitation spectrums in the system with SBO method, this enables us to determine the boundaries of different phases in the magnetic phase diagrams.

The rest of the paper is organized as follows. In Sec.~\ref{section_model}, we give a brief introduction of the spin-1 bosonic system trapped in an optical lattice. In Sec.~\ref{section_MI-SF}, we consider the MI-SF phase transition process with SBO method and analyze the symmetry properties in the insulating and superfluid phases, and we also give the self-consistent SBO phase diagrams in this section. In Sec.~\ref{section_Mott1}, we analyze the magnetic phase diagram in the first ($n=1$) Mott lobe with linear and quadratic Zeeman effect. In Sec.~\ref{section_Mott2}, we analyze the magnetic phase diagram in the second ($n=2$) Mott lobe with linear Zeeman effect. Finally, in Sec.~\ref{section_conlusion}, we summarize our results and give some conclusions. Some of the technical calculation details are presented in the appendix section.

\section{spin-1 model}

\label{section_model}

Spin-1 bosons trapped in a spin-independent optical lattice can be described with Bose-Hubbard Hamiltonian supplemented by spinor interaction in each lattice site~\cite{PhysRevA.68.063602,PhysRevLett.88.163001,RevModPhys.85.1191}. The Hamiltonian is given by
\begin{eqnarray}
\hat{H}&=&-t\sum_{\langle ij \rangle,\sigma}(\hat{a}^\dagger_{i\sigma}\hat{a}_{j\sigma}+\hat{a}^\dagger_{j\sigma}\hat{a}_{i\sigma})+\frac{U_0}{2} \sum_i \hat{n}_i(\hat{n}_i-1)\nonumber\\
&&+\frac{U_2}{2}\sum_i (\vec{S}_i^2-2\hat{n}_i)-\mu\sum_{i}\hat{n}_i,
\label{Hamiltonian}
\end{eqnarray}
where $\hat{a}_{i\sigma}$ is the annihilation operator of one boson in site $i$ with spin-component $\sigma\in\{-1,0,1\}$, $\hat{n}_i=\sum_{\sigma}\hat{a}^\dagger_{i\sigma}\hat{a}_{i\sigma}$, and $\mu$ is the chemical potential that controls the total number of bosons in the system. The total spin in site $i$ is defined as
\begin{eqnarray}
\vec{S}_i&=&\sum_{\sigma\sigma^\prime}\hat{a}^\dagger_{i\sigma}\vec{F}_{\sigma\sigma^\prime}\hat{a}_{i\sigma^\prime},
\label{Spin_i}
\end{eqnarray}
where $\vec{F}_{\sigma\sigma^\prime}$ are the spin operators for spin-1 bosons.

The first term in Eq.~(\ref{Hamiltonian}) is the kinetic energy term that allows particles to hop between neighbour sites, the second term describes spin-independent interaction while the third term describes spin-exchange interaction which is due to the difference in scattering lengths for $S=0$ and $S=2$ channels~\cite{PhysRevLett.81.742}. The spin-independent interaction is always repulsive ($U_0>0$), and $U_2<0$ case (e.g. $^{87}$Rb) is referred to as "ferromagnetic" case whereas the $U_2>0$ case (e.g. $^{23}$Na) is referred to as "antiferromagnetic".

The on-site Hamiltonian in Eq.~(\ref{Hamiltonian}) (neglecting the hopping term) can be described with eigenstate $|S_i,m_i;n_i\rangle$, which represents total spin ($S_i$), spin-projection on z-axis ($m_i$) and total number of bosons ($n_i$) in site $i$~\cite{PhysRevLett.81.5257,PhysRevA.70.043628,PhysRevA.68.043612}. The corresponding eigenenergy is given by
\begin{eqnarray}
E^{(0)}(S_i,n_i)&=&-\mu n_i+\frac{U_0}{2}n_i(n_i-1)\nonumber\\
&&+\frac{U_2}{2}\left[S_i(S_i+1)-2n_i\right].
\label{Energy_onsite}
\end{eqnarray}

The states are degenerate with respect to different $m_i$. For each lattice site, the eigenstates are constrained by $0\leq S_i\leq n_i$, $-S_i\leq m_i\leq S_i$ and $S_i+n_i$ is an even number (due to the symmetry of spin-wave function~\cite{PhysRevA.54.4534}).

In the MI phase, the ground state is determined by minimizing the on-site energy $E^{(0)}(S_i,n_i)$ (Eq.~(\ref{Energy_onsite})). For antiferromagnetic case ($U_2>0$), the ground state in MI phase can be $|0,0;n\rangle$ (for even $n$) or $|1,m_S;n\rangle$ (for odd $n$). While for ferromagnetic case ($U_2<0$), the ground state is $|n,m_S;n\rangle$ for both even and odd filling number $n$.

\section{MI-SF phase transition}

\label{section_MI-SF}

For finite hopping energy $t$, the Mott insulator-superfluid phase transition is induced by change in fluctuation of particle number in individual lattice site. The transition process of spin-1 boson system can be calculated with standard basis operator (SBO) method~\cite{PhysRevB.5.1106,sheshadri1993superfluid,PhysRevA.92.013602}.

We start by expanding Hamiltonian in Eq.~(\ref{Hamiltonian}) on the basis of eigenstates of on-site Hamiltonian (denoted as $|i\mu\rangle$ in expression)
\begin{eqnarray}
\hat{H}&=&\sum_{i\mu}E_{\mu}^i\hat{L}_{\mu\mu}^i-t\sum_{\langle ij\rangle}\sum_{\mu\mu^\prime\nu\nu^\prime}T^{ij}_{\mu\mu^\prime\nu\nu^\prime} \hat{L}^i_{\mu\mu^\prime}\hat{L}^j_{\nu\nu^\prime},
\label{Hamiltonian_expand}
\end{eqnarray}
where $\hat{L}^i_{\mu\mu^\prime}=|i\mu\rangle\langle i\mu^\prime|$ and $T^{ij}_{\mu\mu^\prime\nu\nu^\prime}=\sum_{\sigma}(c^{i\sigma}_{\mu\mu^\prime}d^{j\sigma}_{\nu\nu^\prime} + d^{i\sigma}_{\mu\mu^\prime}c^{j\sigma}_{\nu\nu^\prime})$ with the definitions $c^{i\sigma}_{\mu\mu^\prime}=\langle i\mu|\hat{a}_{i\sigma}|i\mu^\prime\rangle$ and $d^{i\sigma}_{\mu\mu^\prime}=\langle i\mu|\hat{a}^\dagger_{i\sigma}|i\mu^\prime\rangle$. $E^i_{\mu}$ in Eq.~(\ref{Hamiltonian_expand}) is the engenenergy of the eigenstate $|i\mu\rangle$ in site $i$.

Then the retarded Green's function can also be expanded on the basis of $|i\mu\rangle$
\begin{eqnarray}
G^{i\alpha j\beta}(t-t^\prime)&=&-i\Theta(t-t^\prime)\langle [\hat{a}_{i\alpha}(t),\hat{a}^\dagger_{j\beta}(t^\prime)]\rangle\nonumber\\
&=&\sum_{\mu\mu^\prime\nu\nu^\prime}c^{i\alpha}_{\mu\mu^\prime} d^{j\beta}_{\nu\nu^\prime}G^{ij}_{\mu\mu^\prime,\nu\nu^\prime}(t-t^\prime),
\label{Green_function_expand}
\end{eqnarray}
where we have defined $G^{ij}_{\mu\mu^\prime,\nu\nu^\prime}(t-t^\prime)=-i\Theta(t-t^\prime) \langle [\hat{L}^i_{\mu\mu^\prime}(t),\hat{L}^j_{\nu\nu^\prime}(t^\prime)]\rangle$.

For the homogeneous lattice with translational symmetry, we can eliminate site-number and give the motion equation for Green's function (see appendix.A)
\begin{eqnarray}
&&(\omega+E_{\mu}-E_{\mu^\prime})G_{\mu\mu^\prime,\nu\nu^\prime}(\mathbf{k},\omega)-D_{\mu\mu^\prime}\delta_{\mu\nu^\prime}\delta_{\mu^\prime\nu}\nonumber\\
&=&D_{\mu\mu^\prime}\epsilon(\mathbf{k}) \sum_{\xi\xi^\prime}T_{\mu^\prime\mu\xi\xi^\prime}G_{\xi\xi^\prime,\nu\nu^\prime}(\mathbf{k},\omega),
\label{Green_motion}
\end{eqnarray}
where $G_{\mu\mu^\prime,\nu\nu^\prime}(\mathbf{k},\omega)$ is derived from $G^{ij}_{\mu\mu^\prime,\nu\nu^\prime}(t-t^\prime)$ via Fourier transformation, $D_{\mu\mu^\prime}=D_{\mu}-D_{\mu^\prime}$ with definition $D_{\mu}=\langle\hat{L}_{\mu\mu}\rangle$ and $\epsilon(\mathbf{k})=-2t\sum_{s=1}^{\text{dim}}\cos{k_s l}$ ($l$ is the length between neighbour sites in the lattice).

Then the Green's function which describes annihilation of spin-$\alpha$ and construction of spin-$\beta$ bosons can be derived from the Fourier transformation form of Eq.~(\ref{Green_function_expand})
\begin{eqnarray}
G^{\alpha\beta}(\mathbf{k},\omega)&=&\sum_{\mu\mu^\prime\nu\nu^\prime} c_{\mu\mu^\prime}^\alpha d_{\nu\nu^\prime}^\beta G_{\mu\mu^\prime,\nu\nu^\prime}(\mathbf{k},\omega).
\label{Green_alpha_beta}
\end{eqnarray}

In the calculation of $G^{\alpha\beta}(\mathbf{k},\omega)$, we can treat it as matrix element of a $3\times 3$ matrix $\mathbf{G}(\mathbf{k},\omega)$ with the definition $G^{\alpha\beta}=[\mathbf{G}]_{\alpha\beta}$. Then matrix $\mathbf{G}(\mathbf{k},\omega)$ can be solved (in appendix.A) as
\begin{eqnarray}
\mathbf{G}(\mathbf{k},\omega)&=&[\mathbf{I}_{3\times3}-\epsilon(\mathbf{k})\mathbf{\Pi}(\mathbf{k},\omega)]^{-1} \mathbf{\Pi}(\mathbf{k},\omega),\label{Green_matrix}\\
\mathbf{\Pi}(\mathbf{k},\omega)&=&\mathbf{N}_{11}(\omega)+\epsilon(\mathbf{k})\mathbf{\Gamma}(\mathbf{k},\omega),\\
\mathbf{\Gamma}(\mathbf{k},\omega)&=&\mathbf{N}_{12}(\omega)[\mathbf{I}_{3\times3}-\epsilon(\mathbf{k})\mathbf{N}_{22}(\omega)]^{-1}\mathbf{N}_{21}(\omega),
\end{eqnarray}
where $\mathbf{I}_{3\times3}$ represents $3\times 3$ identity matrix and matrices $\mathbf{N}_{mn}$ are defined as
\begin{eqnarray}
&&\begin{pmatrix}\mathbf{N}_{11}^{\alpha,\beta}(\omega) & \mathbf{N}_{12}^{\alpha,\beta}(\omega)\\ \mathbf{N}_{21}^{\alpha,\beta}(\omega) & \mathbf{N}_{22}^{\alpha,\beta}(\omega)\end{pmatrix}\nonumber\\
&=&\sum_{\mu\mu^\prime}\frac{D_{\mu\mu^\prime}}{\omega+E_{\mu}-E_{\mu^\prime}}
\begin{pmatrix}c_{\mu\mu^\prime}^{\alpha}d_{\mu^\prime\mu}^{\beta} & c_{\mu\mu^\prime}^{\alpha}c_{\mu^\prime\mu}^{\beta}\\
d_{\mu\mu^\prime}^{\alpha}d_{\mu^\prime\mu}^{\beta} & d_{\mu\mu^\prime}^{\alpha}c_{\mu^\prime\mu}^{\beta}\end{pmatrix}.
\end{eqnarray}

Usually we have $\mathbf{N}_{12}=\mathbf{N}_{21}=0$, so the Green's function matrix is simplified to be
\begin{eqnarray}
\mathbf{G}(\mathbf{k},\omega)=\frac{\mathbf{N}_{11}(\omega)}{\mathbf{I}_{3\times 3}-\epsilon(\mathbf{k})\mathbf{N}_{11}(\omega)}.
\end{eqnarray}

In the MI-SF phase transition process, we assume the one-particle excitation is described by operator $\hat{A}_i^\dagger=\sum_{\sigma=\pm1,0}\chi_{\sigma}\hat{a}_{i\sigma}^\dagger$ with $\sum_{\sigma}|\chi_\sigma|^2=1$, which indicates that the superfluid order parameter can be expressed as $\psi_{\sigma}=\sqrt{n_s}\chi_{\sigma}$ ($\sigma=0,\pm1$ and $n_s$ is the number of condensate atoms in each site). Then the most probable excitation is determined by maximizing $|\langle\langle[A,A^\dagger]\rangle\rangle|_{\mathbf{k}=\mathbf{0},\omega=0}=\sum_{\sigma_1\sigma_2}\chi_{\sigma_1}^*\chi_{\sigma_2}|G^{\sigma_1\sigma_2}(\mathbf{k} =\mathbf{0},\omega=0)|$, this is equivalent to diagonalizing matrix $\mathbf{G}(\mathbf{k}=\mathbf{0},\omega=0)$ (or matrix $\mathbf{N}_{11}(\omega=0)$) and keeping the eigenvector with maximum absolute eigenvalue. The phase boundary is determined by the poles of the diagonalized Green's function matrix.

\subsection{Ferromagnetic case ($U_2<0$)}

For ferromagnetic case ($U_2<0$), theorem in Ref.~\cite{PhysRevLett.110.130405} shows that the ground state universally exhibits saturated ferromagnetic in superfluid and Mott insulator phases, and the same result is predicted in numerical calculations with mean-field approximation~\cite{PhysRevB.77.014503} and 1D-QMC simulations~\cite{PhysRevLett.102.140402}. In this subsection, we will give detailed analytical analysis of this behavior with SBO method.

The ground state in the MI phase is $|n,m_S;n\rangle$, which is degenerate for $m_s=-n,-n+1,...,n$. Without loss of generality, we assume that system in MI phase is in superposition state $|\psi_0\rangle=\sum_{s=-n}^{n}c_s|n,s;n\rangle$ ($\sum_{s=-n}^n|c_s|^2=1$) with $D_{\psi_0}=1$. For the convenience of following discussions, we also define
\begin{eqnarray}
A^{(\alpha)s,s+\alpha}_{n,n\pm1}&=&\langle n\pm1,s+\alpha;n+1|\hat{a}_{\alpha}^\dagger|n,s;n\rangle,\\
B^{(\alpha)s,s-\alpha}_{n,n-1}&=&\langle n-1,s-\alpha;n-1|\hat{a}_{\alpha}|n,s;n\rangle,
\end{eqnarray}
and
\begin{eqnarray}
\Omega_1&=&\frac{1}{E^{(0)}(n-1,n+1)-E^{(0)}(n,n)},\\
\Omega_2&=&\frac{1}{E^{(0)}(n+1,n+1)-E^{(0)}(n,n)},\\
\Omega_3&=&\frac{1}{E^{(0)}(n-1,n-1)-E^{(0)}(n,n)},
\end{eqnarray}
then matrix $\mathbf{N}_{11}(0)$ can be derived
\begin{eqnarray}
\mathbf{N}_{11}^{0,0}&=&\sum_s|c_s|^2\Big[\Omega_1(A^{(0)s,s}_{n,n-1})^2 + \Omega_2(A^{(0)s,s}_{n,n+1})^2\nonumber\\
&&+ \Omega_3(B^{(0)s,s}_{n,n-1})^2\Big],
\end{eqnarray}
\begin{eqnarray}
\mathbf{N}_{11}^{1,1}&=&\sum_s|c_s|^2\Big[\Omega_1(A^{(1)s,s+1}_{n,n-1})^2 +\Omega_2(A^{(1)s,s+1}_{n,n+1})^2\nonumber\\
&&+\Omega_3(B^{(1)s,s-1}_{n,n-1})^2\Big],
\end{eqnarray}
\begin{eqnarray}
\mathbf{N}_{11}^{-1,-1}&=&\sum_s|c_s|^2\Big[\Omega_1(A^{(-1)s,s-1}_{n,n-1})^2 +\Omega_2(A^{(-1)s,s-1}_{n,n+1})^2\nonumber\\
&&+\Omega_3(B^{(-1)s,s+1}_{n,n-1})^2\Big],
\end{eqnarray}
\begin{eqnarray}
&&\mathbf{N}_{11}^{0,1}=\sum_s c_sc_{s+1}^*\Big[\Omega_1A^{(1)s,s+1}_{n,n-1} A^{(0)s+1,s+1}_{n,n-1} \nonumber\\ &&+\Omega_2A^{(1)s,s+1}_{n,n+1}A^{(0)s+1,s+1}_{n,n+1}
+\Omega_3B^{(1)s+1,s}_{n,n-1}B^{(0)s,s}_{n,n-1}\Big],\nonumber\\
\end{eqnarray}
\begin{eqnarray}
&&\mathbf{N}_{11}^{0,-1}=\sum_s c_sc_{s-1}^*\Big[\Omega_1A^{(-1)s,s-1}_{n,n-1} A^{(0)s-1,s-1}_{n,n-1} \nonumber\\ &&+\Omega_2A^{(-1)s,s-1}_{n,n+1}A^{(0)s-1,s-1}_{n,n+1}
+\Omega_3B^{(-1)s-1,s}_{n,n-1}B^{(0)s,s}_{n,n-1}\Big],\nonumber\\
\end{eqnarray}
\begin{eqnarray}
&&\mathbf{N}_{11}^{1,-1}=\sum_s c_{s+1}c_{s-1}^*\Big[\Omega_1A^{(-1)s+1,s}_{n,n-1} A^{(1)s-1,s}_{n,n-1} \nonumber\\ &&+\Omega_2A^{(-1)s+1,s}_{n,n+1}A^{(1)s-1,s}_{n,n+1}
+\Omega_3B^{(-1)s-1,s}_{n,n-1}B^{(1)s+1,s}_{n,n-1}\Big],\nonumber\\
\end{eqnarray}
and
\begin{eqnarray}
\mathbf{N}_{11}^{1,0}&=&(\mathbf{N}_{11}^{0,1})^*,\\
\mathbf{N}_{11}^{-1,0}&=&(\mathbf{N}_{11}^{0,-1})^*,\\
\mathbf{N}_{11}^{-1,1}&=&(\mathbf{N}_{11}^{1,-1})^*.
\end{eqnarray}

Diagonalizing the matrix $\mathbf{N}_{11}$ directly is a tough work, especially when the filling number $n$ is very large. So here we come up with a more effective method to solve the problem.

When we set $c_n=1$ (or $c_{-n}=1$) in the ground state $|\psi_0\rangle$, matrix $\mathbf{N}_{11}$ is diagonal with maximum eigenvalue $\lambda_m=\Omega_2(A^{(1)n,n+1}_{n,n+1})^2+\Omega_3(B^{(1)n,n-1}_{n,n-1})^2=(n+1)\Omega_2+n\Omega_3$. Then we prove that the matrix $\lambda_m\mathbf{I}_{3\times3}-\mathbf{N}_{11}$ can be expanded as summation of a series of positive-semidefinite matrices, the details of the provement are presented in appendix.B. Thus the eigenvalues $\lambda$ of matrix $\mathbf{N}_{11}$ must satisfy the condition $\lambda\leq\lambda_m$, and the conditions for $\lambda=\lambda_m$ are given as
\begin{eqnarray}
\frac{P^{(\alpha_1)}_B(s)}{P^{(\alpha_2)}_B(s)}&=&R_{\alpha_1,\alpha_2},
\end{eqnarray}
\begin{eqnarray}
R_{1,0}R_{-1,0}=\frac{1}{2},
\end{eqnarray}
where $R_{\alpha_1,\alpha_2}$ is s-independent and we have defined
\begin{eqnarray}
P^{(\alpha)}_A(s)&=&c_{s-\alpha}A^{(\alpha)s-\alpha,s}_{n,n-1},\\
P^{(\alpha)}_B(s)&=&c_{s+\alpha}B^{(\alpha)s+\alpha,s}_{n,n-1}.
\label{P_B_definition}
\end{eqnarray}

In the ground state $|\psi_0\rangle$ (Mott-insulator phase), we have $\langle\vec{S}\rangle^2=n^2$ (see appendix.B), so the MI phase is in ferromagnetic state. Meanwhile the excitation described with the eigenvector $(1,R_{1,0},R_{-1,0})$ indicates that the superfluid phase involved in the phase transition process is also in ferromagnetic state (because $1-2R_{1,0}R_{-1,0}=0$). Therefore the system is ferromagnetic in both superfluid and Mott-insulator phases.

The phase boundary for ferromagnetic case is given by
\begin{eqnarray}
zt_c[(n+1)\Omega_2+n\Omega_3]=1.
\end{eqnarray}

\subsection{Antiferromagnetic case ($U_2>0$)}

For antiferromagnetic case ($U_2>0$), perturbative mean-field approximation (PMFA)~\cite{PhysRevA.70.043628}, density-matrix renormalization group (DMRG)~\cite{PhysRevLett.95.240404} and 1D-QMC~\cite{PhysRevLett.102.140402,PhysRevA.74.035601} studies have shown that the MI phase for even filling lobe is considerably more stable against the superfluid phase than that for odd filling lobe. The MI phase for both even and odd filling lobes is transited to polar superfluid phase with zero global magnetization. In this subsection, we use SBO method to analytically analyze the MI-SF transition process for antiferromagnetic case, of which the ground state energy in the MI phase is minimized by taking minimum total spin $S_i$ in Eq.~(\ref{Energy_onsite}).

For odd filling number case, the ground state is $|1,m_S;n\rangle$, which is degenerate for $m_S=\pm1,0$, the system in MI phase can be in any superposition states of $|1,m_S;n\rangle$. Similarly, we assume the system is in state $|\psi_0\rangle=\sum_{s=\pm1,0} c_{s}|1,s;n\rangle$ ($\sum_{s=\pm1,0}|c_s|^2=1$) with $D_{\psi_0}=1$, then we calculate the corresponding matrix $\mathbf{N}_{11}(0)$ (see appendix.B). Diagonalize matrix $\mathbf{N}_{11}(0)$ and the resulting eigenvalues are
\begin{eqnarray}
&&\lambda_0=3(\Upsilon_2+\Upsilon_4),\\
&&\lambda_{\pm}=\frac{1}{2}(\Upsilon_1+\Upsilon_3+7\Upsilon_2+7\Upsilon_4)\nonumber\\
&&\pm\frac{1}{2}\sqrt{(K_1-K_2)^2+4K_1K_2|c_0^2-2c_1c_{-1}|^2},
\end{eqnarray}
where $K_1=3\Upsilon_2+\Upsilon_3-2\Upsilon_4$ and $K_2=\Upsilon_1-2\Upsilon_2+3\Upsilon_4$ with $\Upsilon_m$ defined as
\begin{eqnarray}
\Upsilon_1&=&\frac{n+1}{3}\frac{1}{E^{(0)}(0,n+1)-E^{(0)}(1,n)},\\
\Upsilon_2&=&\frac{n+4}{15}\frac{1}{E^{(0)}(2,n+1)-E^{(0)}(1,n)},\\
\Upsilon_3&=&\frac{n+2}{3}\frac{1}{E^{(0)}(0,n-1)-E^{(0)}(1,n)},\\
\Upsilon_4&=&\frac{n-1}{15}\frac{1}{E^{(0)}(2,n-1)-E^{(0)}(1,n)}.
\end{eqnarray}

The maximum eigenvalue can be obtained by choosing $\lambda_{+}$ and take the maximum value of $|c_0^2-2c_{1}c_{-1}|^2_{\text{max}}=1$, the corresponding eigenvector $(\chi_0,\chi_1,\chi_{-1})$, which describes the symmetry of superfluid order in the excitation, satisfies the condition $|\chi_0^2-2\chi_1\chi_{-1}|^2=1$. This result indicates that the superfluid phase is in polar state, which is in consistent with perturbation mean-field approximation (PMFA) method~\cite{PhysRevA.70.043628}. Then obviously, the phase boundary for the odd filling number case is given by
\begin{eqnarray}
zt_c\left(\Upsilon_1+4\Upsilon_2+\Upsilon_3+4\Upsilon_4\right)=1.
\end{eqnarray}

For even filling number case, the ground state is $|0,0;n\rangle$, so we can set $D_{|\mu\rangle=|0,0;n\rangle}=1$ if fluctuations are neglected. Then it can be directly derived that matrix $\mathbf{N}_{11}(0)$ is diagonal with $\mathbf{N}_{11}^{\alpha,\beta}=N_0\delta_{\alpha\beta}$, the excitation spectrum is degenerate for spin-$\sigma$ ($\sigma=\pm1,0$) excitations. The phase boundary is derived by setting $1+zt_cN_0=0$ and have
\begin{eqnarray}
zt_c\left(\frac{n/3+1}{\Delta E_1}+\frac{n/3}{\Delta E_2}\right)=1,
\end{eqnarray}
where
\begin{eqnarray}
\Delta E_1&=&E^{(0)}(1,n+1)-E^{(0)}(0,n),\\
\Delta E_2&=&E^{(0)}(1,n-1)-E^{(0)}(0,n).
\end{eqnarray}

To decide the symmetry of superfluid order parameter in the MI-SF phase transition process, the PMFA method has taken fourth-order perturbation into account~\cite{PhysRevA.70.043628}. In the SBO method, we can consider the two-particle excitation, which is described with operator $(\hat{A}^\dagger_i)^2=\sum_{\sigma_1\sigma_2}\chi_{\sigma_1}\chi_{\sigma_2}\hat{a}_{i\sigma_1}^\dagger \hat{a}_{i\sigma_2}^\dagger$. Similar approach can be applied to the two-particle excitation process and gives
\begin{eqnarray}
G^{ij}_{\hat{A}^2}(t-t^\prime)&=&-i\Theta(t-t^\prime)\langle[\hat{A}_i(t)^2,(\hat{A}_j^\dagger(t^\prime))^2]\rangle\nonumber\\
&=&\sum_{\mu\mu^\prime\nu\nu^\prime}G^{ij}_{\mu\mu^\prime,\nu\nu^\prime}(t-t^\prime)c^{\hat{A}^2}_{\mu\mu^\prime}d^{\hat{A}^2}_{\nu\nu^\prime},
\end{eqnarray}
where $c^{\hat{A}^2}_{\mu\mu^\prime}=\langle\mu|\hat{A}^2|\mu^\prime\rangle$ and $d^{\hat{A}^2}_{\nu\nu^\prime}=\langle\nu|(\hat{A}^\dagger)^2|\nu^\prime\rangle$. By Taking the conditions $D_{|\mu\rangle=|0,0;n\rangle}=1$ and $D_{|\mu\rangle\neq|0,0;n\rangle}=0$ as well as the expression in Eq.~(\ref{Green_motion}), the problem is equivalent to maximising absolute value of $G_{\hat{A}^2}(\mathbf{k}=0,\omega=0)$, which can be calculated as
\begin{eqnarray}
&&|G_{\hat{A}^2}(\mathbf{k}=\mathbf{0},\omega=0)|=2(\Lambda_1+\Lambda_3)\nonumber\\
&&+(\Lambda_2+\Lambda_4-\frac{2}{3}\Lambda_1-\frac{2}{3}\Lambda_3)|\chi_0^2-2\chi_1\chi_{-1}|^2,\nonumber\\
\end{eqnarray}
where
\begin{eqnarray}
\Lambda_1&=&\frac{n(n-2)}{15}\frac{1}{E^{(0)}(2,n-2)-E^{(0)}(0,n)},\\
\Lambda_2&=&\frac{n(n+1)}{9}\frac{1}{E^{(0)}(0,n-2)-E^{(0)}(0,n)},\\
\Lambda_3&=&\frac{(n+3)(n+5)}{15}\frac{1}{E^{(0)}(2,n+2)-E^{(0)}(0,n)},\\
\Lambda_4&=&\frac{(n+2)(n+3)}{9}\frac{1}{E^{(0)}(0,n+2)-E^{(0)}(0,n)}.
\end{eqnarray}

It is clear that $\Lambda_2+\Lambda_4-\frac{2}{3}\Lambda_1-\frac{2}{3}\Lambda_3$ is positive, then we need to take the maximum value of $|\chi_0^2-2\chi_1\chi_{-1}|_{\text{max}}=1$, which indicates that the superfluid phase involved in the excitation is in polar state~\cite{PhysRevA.70.043628}. In physics, we can understand the favorable of polar state in the phase transition process is due to the hopping of singlet pairs. The forming of such two-atom dimers in the system can greatly reduce one-particle excitation in even filling number case, even for ferromagnetic interaction ($U_2<0$). This phenomenon is obvious in the SBO phase diagrams shown in the next subsection.

\subsection{Phase diagrams of self-consistent SBO method}

The phase diagrams of MI-SF transition in spin-1 bosonic system have been studied with several approaches, including mean-field approximation~\cite{PhysRevA.70.043628,PhysRevLett.94.110403,PhysRevB.77.014503,PhysRevA.68.043612}, QMC~\cite{PhysRevLett.102.140402,PhysRevB.88.104509}, DMRG~\cite{PhysRevLett.95.240404} and strong-coupling expansion method~\cite{PhysRevA.87.043624}. In this subsection, we will give the phase diagram with self-consistent SBO approach.

In the above two subsections, we have assumed that all atoms stay in the same ground state in MI phase. However, fluctuations exist in the phase diagram even at zero temperature, especially near the phase boundary. The fluctuation term can be calculated with self-consistent SBO approach~\cite{PhysRevA.92.013602}, which uses the spectral theorem~\cite{PhysRevB.5.1106}
\begin{eqnarray}
\langle \hat{L}^j_{\nu\nu^\prime}(t)\hat{L}^i_{\mu\mu^\prime}\rangle&=&\int_{-\infty}^{\infty}\frac{d\omega}{2\pi i}e^{i\omega t}f(\omega)\big[G^{ij}_{\mu\mu^\prime,\nu\nu^\prime}(\omega-i0^+)\nonumber\\
&&-G^{ij}_{\mu\mu^\prime,\nu\nu^\prime}(\omega+i0^+)\big],
\end{eqnarray}
where $f(\omega)=1/(e^{\omega/T}-1)$ is the Bose-statistical function (with $k_B=1$).

Let's take the antiferromagnetic case with odd filling number as an example, for ground state $|\mu_0\rangle=|1,0;n\rangle$ and excited state $|\mu\rangle=|0,0;n-1\rangle$, we apply similar method introduced in spinless model~\cite{PhysRevA.92.013602} and have
\begin{eqnarray}
D_{\mu}&=&\frac{1}{N_s}\sum_{\mathbf{k}}\int_{-\infty}^{\infty}\frac{d\omega}{2\pi i}f(\omega)\Big[F(\mathbf{k},\omega-i0^+)\nonumber\\
&&-F(\mathbf{k},\omega+i0^+)\Big],\\
F(\mathbf{k},\omega)&=&G_{\mu_0\mu,\mu\mu_0}(\mathbf{k},\omega)+\sqrt{\frac{n+1}{n+2}}G_{\mu_1\mu_0,\mu\mu_0}(\mathbf{k},\omega)\nonumber\\
&&+2\sqrt{\frac{n+4}{5(n+2)}}G_{\mu_2\mu_0,\mu\mu_0}(\mathbf{k},\omega),
\end{eqnarray}
where $|\mu_1\rangle=|0,0;n+1\rangle$ and $|\mu_2\rangle=|2,0;n+1\rangle$, $N_s$ is the total number of lattice sites.

The Green's functions in the integration can also be found in the motion equation in Eq.~(\ref{Green_motion}). Then the combined self-consistent equations can be solved afterwards.

\begin{figure}[htb]
\includegraphics[width=\columnwidth]{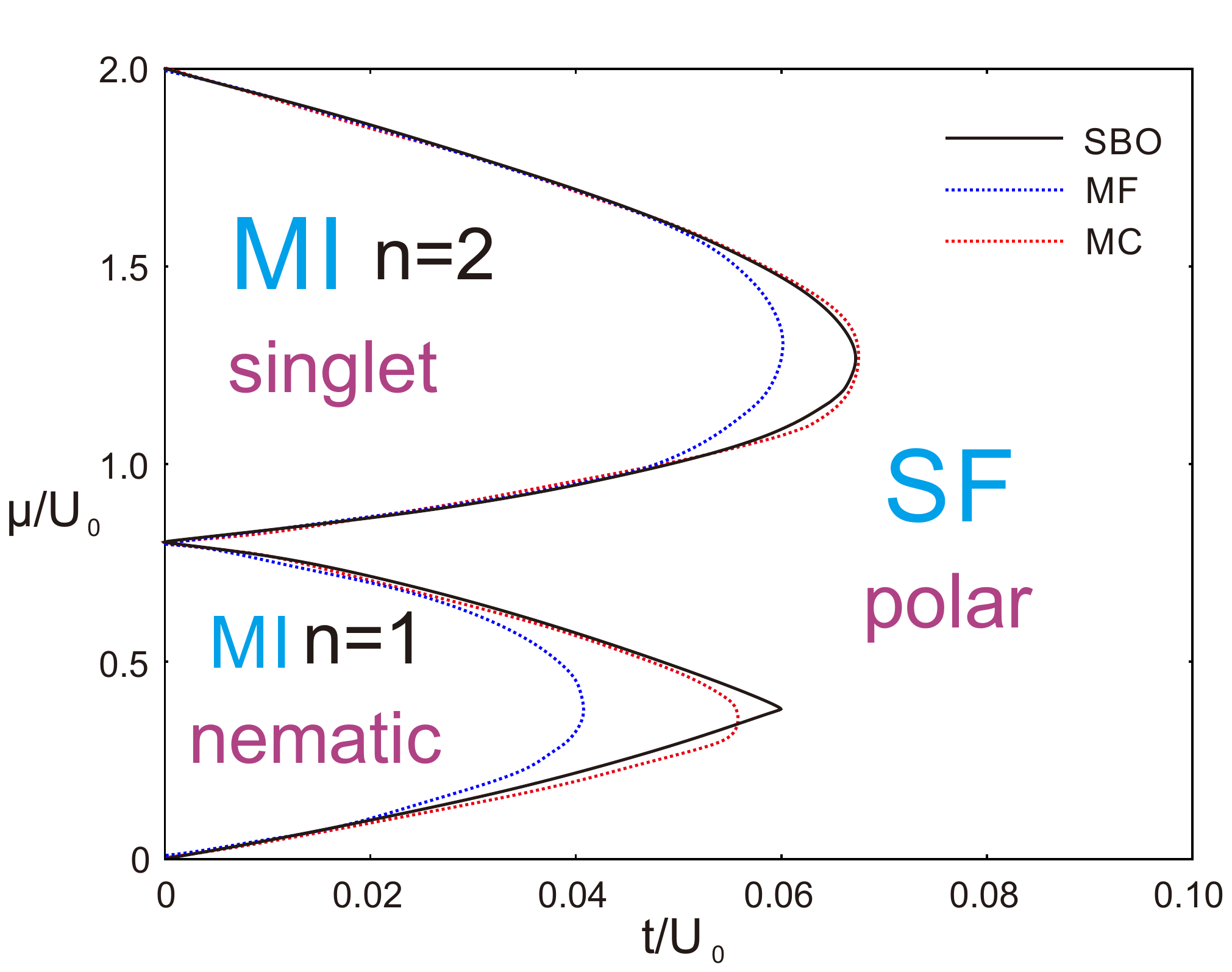}
\caption{Zero-temperature phase diagram for 2-d antiferromagnetic ($U_2=0.1U_0$) case. The mean-field (MF) and Monte-Carlo (MC) results are obtained from Ref.~\cite{PhysRevB.77.014503,PhysRevB.88.104509}.}
\label{figure1}
\end{figure}

In Fig.~\ref{figure1}, we give the phase diagrams for antiferromagnetic case with several methods. Compared to the spinless boson system, we find that the even lobes grow at the expense of the odd lobes, which would entirely disappear when $\frac{U_2}{U_0}\geq0.5$. And for the even lobes, atoms in each site tend to form singlet pairs, so the fluctuations in states $|1,m_S;n\pm1\rangle$ (for even $n$) have been neglected in the SBO phase diagram.

In this figure, we find that the self-consistent SBO zero-temperature phase diagram is in quantitative agreement with quantum Monte-Carlo simulation result~\cite{PhysRevB.88.104509}, which indicates better description of quantum fluctuation in SBO method than in mean-field approximation~\cite{PhysRevA.70.043628,PhysRevB.77.014503}.

\begin{figure}[htb]
\includegraphics[width=\columnwidth]{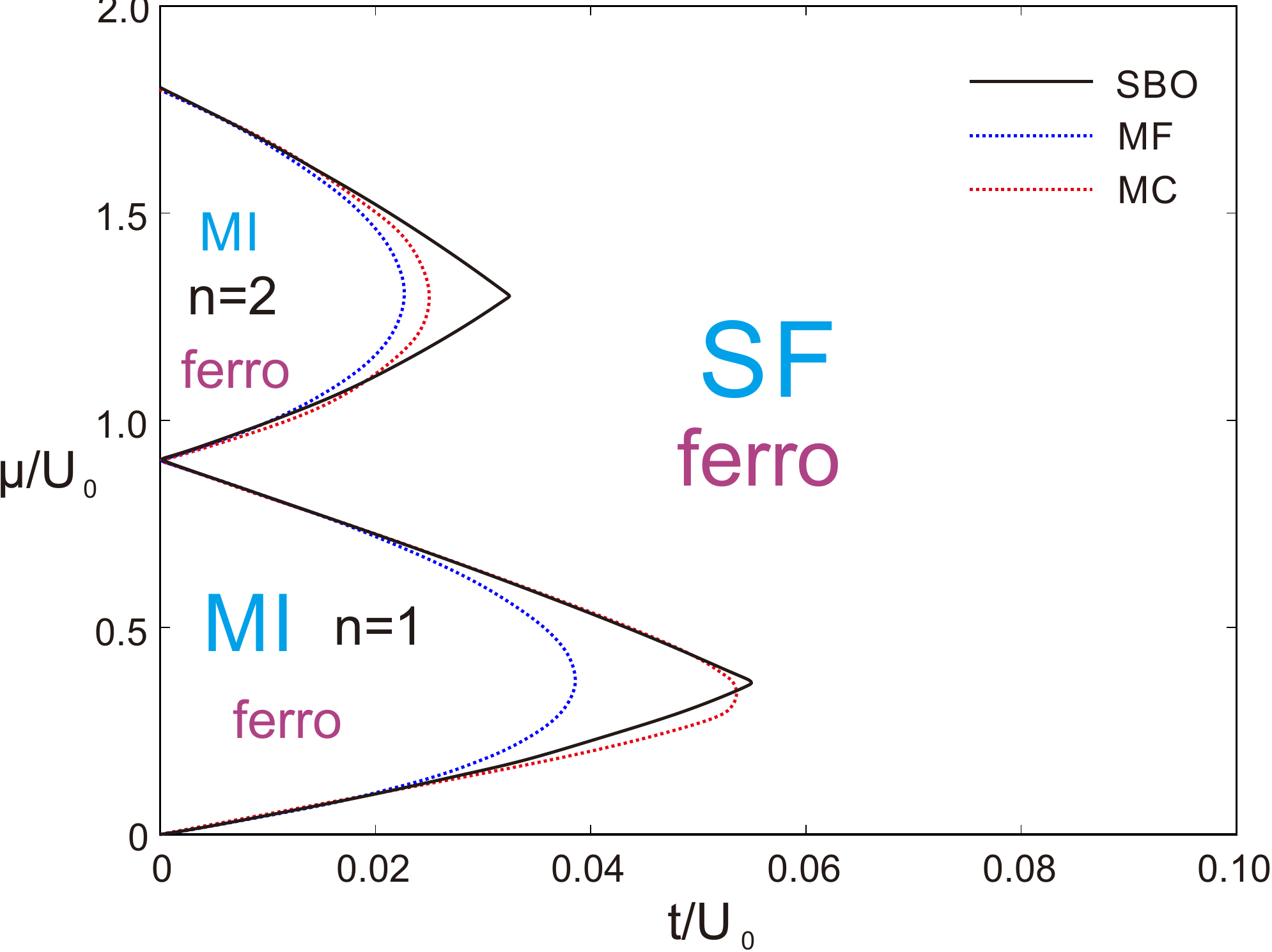}
\caption{Zero-temperature phase diagram for 2-d ferromagnetic ($U_2=-0.1U_0$) case.}
\label{figure2}
\end{figure}

For ferromagnetic case, we have pointed out that the system universally exhibits ferromagnetism in both MI and SF phases. The phase diagrams derived with self-consistent SBO, mean-field approximation~\cite{PhysRevB.77.014503} and Monte-Carlo~\cite{PhysRevB.88.104509} methods are presented in Fig.~\ref{figure2}. Similarly, we can conclude that the SBO method gives more precise phase diagram than mean-field approximation method. For even lobes, the difference between self-consistent SBO and Monte-Carlo results can be explained as due to dimer-effect in the system (consider two-particle excitation), which tends to decrease quantum fluctuations in states $|S,m_S;n\pm1\rangle$ (for even $n$) and induces lower critical hopping energy in the phase diagram.

If external magnetic field is applied to the system, the Hamiltonian is changed by linear Zeeman effect with $\hat{H}^\prime=\hat{H}-\eta\sum_i \hat{S}_i^z$. The external field-induced term tends to change both Mott-insulator and superfluid phases in the phase transition process~\cite{PhysRevA.68.043612}. The result can be derived with similar procedures introduced for non-external-field case.

\begin{figure}[htb]
\includegraphics[width=\columnwidth]{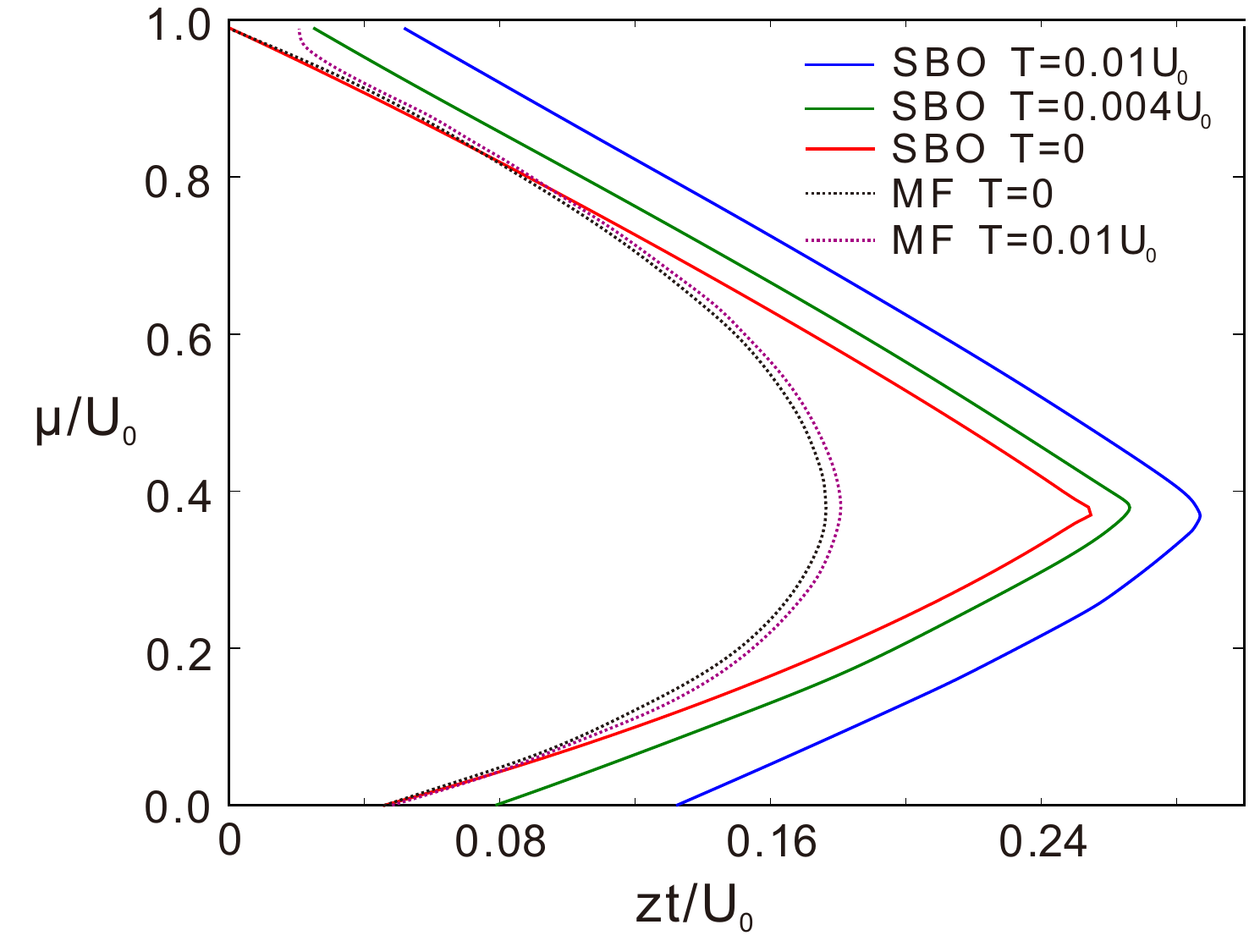}
\caption{Finite temperature phase diagrams for single-atom filling antiferromagnetic system in external field. Parameters are chosen as $U_2=0.04U_0$ and $\eta=0.05U_0$.}
\label{figure3}
\end{figure}

At finite temperature, besides the states involved in the excitations from ground state, we need to further take the fluctuations in other states into consideration. For example, for the single-atom filling ($n=1$) case of antiferromagnetic system with external field, the ground state is $|1,1;1\rangle$, spin-1 excitation takes place in the MI-SF phase transition process. Then besides the states $|0,0;0\rangle$ and $|2,2;2\rangle$, which are involved in the spin-1 excitation from $|1,1;1\rangle$, we need further consider the fluctuations in states $|1,\alpha;1\rangle$ ($\alpha=-1,0$) and $|0,0;2\rangle$, whereas other states are neglected because the energy differences are much larger than temperature ($k_B=1$). In Fig.~\ref{figure3}, we give the phase diagrams for finite temperature system with self-consistent SBO method and mean-field approximation~\cite{ohliger2008thermodynamic}. By comparison, we find that the SBO results are much more sensitive to the change of temperature than mean-field results. Because when temperature increases but is still low enough, the fluctuations in the spin-1 excitation-involved states can be enhanced much greater than those in other states in consideration.

\section{insulating state with one atom per site}

\label{section_Mott1}

The particle excitation properties of the spin-1 bosonic system are shown in Fig.~\ref{figure1} and Fig.~\ref{figure2}. However, the inner spin properties in the Mott lobes have also attracted many attentions, even before the study of cold atom system~\cite{PhysRevB.7.4267}.

In second-order perturbation theory in $t$, the system in the Mott lobe with one atom per site is given by super-exchange processes, which can be described with an effective bilinear-biquadratic spin Hamiltonian~\cite{PhysRevA.68.063602,PhysRevLett.90.250402,PhysRevLett.106.105302,PhysRevA.80.053615}
\begin{eqnarray}
\hat{H}_{\text{eff}}&=&-\sum_{\langle ij\rangle}\left[J_1\vec{S}_i\cdot\vec{S}_j + J_2(\vec{S}_i\cdot\vec{S}_j)^2\right]\nonumber\\
&&+\sum_i\left(\hat{H}_i^L+\hat{H}_i^Q\right).
\label{Hamiltonian_eff_1}
\end{eqnarray}

The absence of higher order terms, such as $(\vec{S}_i\cdot\vec{S}_j)^3$, is due to the fact that the product of any three spin operators for an $S=1$ particle can be expressed via lower order terms. The parameters in the Hamiltonian can be calculated with perturbation theory, which gives $J_1=\frac{2t^2}{g_2}$ and $J_2=\frac{4t^2}{3g_0}+\frac{2t^2}{3g_2}$ with $g_{0,2}=4\pi\hbar^2 a_{0,2}/m_a$ representing the scattering strengths of spin-0 and spin-2 channels. In the cold-atom system, we assume both of them are positive. For antiferromagnetic case ($a_2>a_0$), $J_2>J_1$, and for ferromagnetic case ($a_0>a_2$), $J_1>J_2$. $\hat{H}_i^L$ and $\hat{H}_i^Q$ in Eq.~(\ref{Hamiltonian_eff_1}) represent linear Zeeman effect (LZE) and quadratic Zeeman effect (QZE) respectively when a magnetic field is applied to the system,
\begin{eqnarray}
\hat{H}_i^L&=&-\lambda \hat{S}_i^z,\\
\hat{H}_i^Q&=&q(\hat{S}_i^z)^2.
\end{eqnarray}
In the linear term we have included the Lagrangian multiplier due to magnetization conservation~\cite{stenger1999spin}. In our following discussion, we assume $q>0$, and the result is symmetrical for $\pm\lambda$.

We find the Hamiltonian in Eq.~(\ref{Hamiltonian_eff_1}) is similar to that of spin-1 Heisenberg model with biquadratic exchange, which has been studied in Ref.~\cite{PhysRevB.7.4267,Micnas1976Spin,Papanicolaou1988Unusual}. Then to solve the system with standard basis operator (SBO) method, we expand the Hamiltonian in Eq.~(\ref{Hamiltonian_eff_1}) on the basis $|m\rangle=|1,m;1\rangle$ and obtain the expression in appendix A (Eq.~(\ref{Hamiltonian_form})) with
\begin{eqnarray}
V^i_{mm^\prime}&=&\delta_{mm^\prime}V_{m}=\delta_{mm^\prime}(-\lambda m+qm^2),
\label{E_one}
\end{eqnarray}
and
\begin{eqnarray}
T^{ij}_{mm^\prime nn^\prime}&=&-J_1\langle m|_i\langle n|_j\vec{S}_i\cdot\vec{S}_j|m^\prime\rangle_i |n^\prime\rangle_j\nonumber\\
&&-J_2 \langle m|_i\langle n|_j(\vec{S}_i\cdot\vec{S}_j)^2|m^\prime\rangle_i |n^\prime\rangle_j.
\label{T_one}
\end{eqnarray}

Substitute the expressions in Eq.~(\ref{E_one}) and Eq.~(\ref{T_one}) into the motion equation in the appendix (Eq.~(\ref{Green_motion2})), we derive
\begin{eqnarray}
&&(\omega+E_m-E_{m^\prime})G_{mm^\prime,nn^\prime}(\mathbf{k},\omega)-\delta_{mn^\prime}\delta_{nm^\prime}D_{mm^\prime}\nonumber\\
&=&-\zeta(\mathbf{k}) D_{mm^\prime}\sum_{rr^\prime}T_{m^\prime m rr^\prime}G_{rr^\prime,nn^\prime}(\mathbf{k},\omega),
\label{motion_one}
\end{eqnarray}
where we have defined $\zeta(\mathbf{k})=-2\sum_{s=1}^{\text{dim}}\cos{k_sl}$ and $E_m=V_m+z\sum_{n}D_n T_{mm nn}$.

In our system, the lattice is initially deep enough that the hopping energy can be neglected ($t=J_1=J_2=0$), so the ground state, which can be $|0\rangle$ (nematic) or $|\pm1\rangle$ (ferromagnetic), is determined by $\lambda$ and $q$, and the boundary is given by $q=|\lambda|$. Then increase the hopping energy by reducing the lattice depth, we have $J_1>0$ and $J_2>0$, the system will be excited from ground state, the spin-excitation can be described with SBO method.

When $\lambda>0$ and $\lambda<q$, the ground state is $|0\rangle$, and it will be excited to state $|1\rangle$ first, this process is described with Green's function $G_{01,nn^\prime}(\mathbf{k},\omega)$. Based on the motion equation in Eq.~(\ref{motion_one}), we can write the combined equations related to $G_{01,10}(\mathbf{k},\omega)$
\begin{eqnarray}
\mathbf{M}\begin{pmatrix} G_{01,10}(\mathbf{k},\omega) \\ G_{-10,10}(\mathbf{k},\omega)\end{pmatrix}=\begin{pmatrix}D_{01} \\ 0\end{pmatrix},
\end{eqnarray}
with
\begin{eqnarray}
\mathbf{M}&=&\begin{pmatrix} -J_1D_{01}\zeta(\mathbf{k}) & (J_2-J_1)D_{01}\zeta(\mathbf{k}) \\ (J_2-J_1)D_{-10}\zeta(\mathbf{k}) & -J_1D_{-10}\zeta(\mathbf{k})\end{pmatrix}\nonumber\\
&&+\begin{pmatrix}\omega+\Delta E_1 & 0\\ 0 & \omega+\Delta E_2\end{pmatrix},
\end{eqnarray}
where $\Delta E_1=E_0-E_1$ and $\Delta E_2=E_{-1}-E_0$.

The excitation spectrum is determined by the solution of equation $\det{\mathbf{M}}=0$. If we neglect the fluctuations and assume $D_0=1$ and $D_{\pm1}=0$, the equation corresponds to
\begin{eqnarray}
&&(\omega+\lambda)^2\nonumber\\
&=&(q+zJ_2+J_1\zeta(\mathbf{k}))^2-(J_2-J_1)^2\zeta(\mathbf{k})^2.
\label{det_1}
\end{eqnarray}

We find that in Eq.~(\ref{det_1}), if $q$ is sufficiently large, the solution of $\omega$ in the equation is always non-zero, leading to positive excitation gap in the excitation, then the phase transition in the system is denied and the the system is stable in state $|0\rangle$ (nematic state). To determine the boundary, we analyze the minimum value of the expression on the right side of Eq.~(\ref{det_1}) (see appendix.C).

For antiferromagnetic case ($J_2>J_1$), we have $q>0>zJ_2-z\frac{J_2^2}{J_1}$, so the minimum (we label it as $\text{Min}$) is located at the point $\mathbf{k}=\mathbf{0}$. If $\lambda^2<\text{Min}$, the excitation gap turns to be positive, thus the excitation is denied. So the boundary is given by
\begin{eqnarray}
\lambda^2=(q+zJ_2-zJ_1)^2-z^2(J_2-J_1)^2.
\label{bound_one}
\end{eqnarray}

\begin{figure}[htb]
\includegraphics[width=\columnwidth]{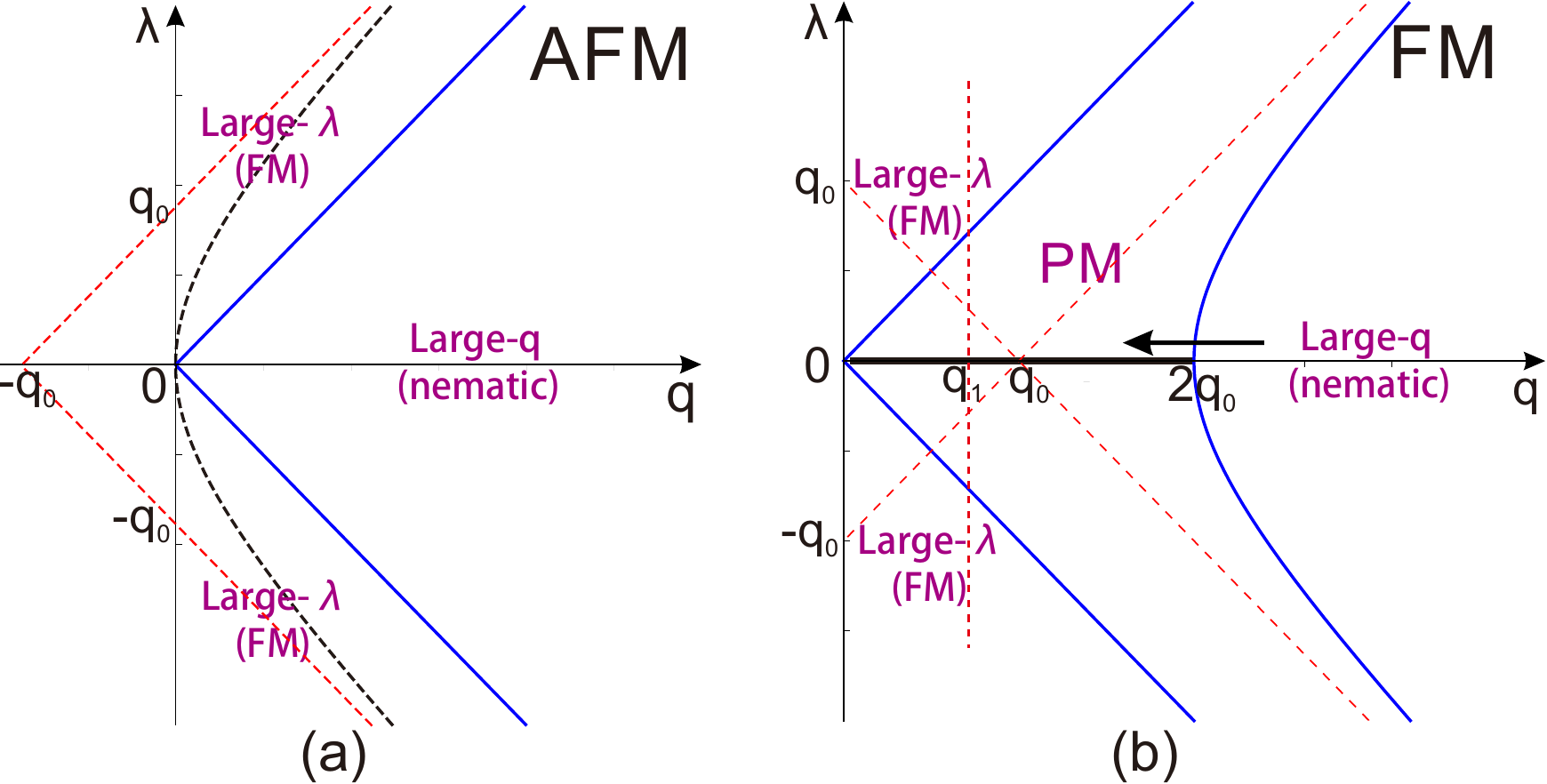}
\caption{Magnetic phase diagrams in $n=1$ Mott lobe for antiferromagnetic case (a) and ferromagnetic case (b). The bold black line on the $q$-axis represents XY-FM phase~\cite{PhysRevLett.106.105302}, and the black arrow along $-q$ direction indicates the change of the relative location for $q=Q_0$ when we increase the hopping energy $t$ from zero to finite. In the figure we have defined $q_0=z|J_2-J_1|$ and $q_1=z(J_2-J_2^2/J_1)$. We note that $q_1<q_0$ in the ferromagnetic case ($J_1>J_2$).}
\label{figure4}
\end{figure}

While for ferromagnetic case ($J_1>J_2$), if $q\geq zJ_2-z\frac{J_2^2}{J_1}$, the minimum is located at the point $\mathbf{k}=\mathbf{0}$, and the phase boundary is also given by Eq.~(\ref{bound_one}). If $0<q<zJ_2-z\frac{J_2^2}{J_1}$, the minimum seems to be located at the global minimum point with $\mathbf{k}\neq\mathbf{0}$. However, we note that, for any $q=Q_0>0$, when we increase hopping energy $t$ from 0 to positive, it will always pass the point $2q_0$ before reaching $q_1$ (see Fig.~\ref{figure4}(b)), so the excitations happen before $q<zJ_2-zJ_2^2/J_1$, we can still take $\mathbf{k}=0$ in calculating the excitation gap. So in this case, the boundary of large-$q$ (nematic) phase is still given by Eq.~(\ref{bound_one}).

Similarly, when $\lambda>q>0$, we start from $|1\rangle$ and consider the excitation to $|0\rangle$ (excite to $|0\rangle$ first), the excitation spectrum can be solved as $\omega_{\mathbf{k}}=\lambda-q+zJ_1+J_1\zeta(\mathbf{k})$, which is always positive.

In Fig.~\ref{figure4}, we divide the $\lambda$-$q$ plane (with solid blue curves) into several regimes, corresponding to several different phases. Each phase represents the ground state and excitation properties in the regime when we increase hopping energy $t$ from $0$ to finite. For example, the nematic (ferromagnetic) phase indicates that the ground state is $|0\rangle$ ($|\pm1\rangle$) and the spin excitation is denied when $t$ increases, while the partially magnetic phase indicates that the system initially in state $|0\rangle$ can be excited to state $|\pm1\rangle$ when $t$ increases.

We can take a look at the results in some special cases. For example, when $\lambda=q=0$, states $|\pm1\rangle$ are symmetric by rotation, so we have $D_1=D_{-1}$. Then we can solve the excitation spectrum from $|0\rangle$ to $|\pm1\rangle$ as
\begin{eqnarray}
\omega_{\mathbf{k}}=D_{01}\sqrt{(zJ_2+J_1\zeta(\mathbf{k}))^2-(J_2-J_1)^2\zeta(\mathbf{k})^2},
\label{spectrum_lambda0_q0}
\end{eqnarray}
the result in Eq.~(\ref{spectrum_lambda0_q0}) is the same as that derived with systematic $1/n$ expansion method~\cite{Papanicolaou1988Unusual} besides an additional fluctuation-correction term $D_{01}$.

When $\lambda=0$ and $q>0$, states $|\pm1\rangle$ are still symmetric by rotation, and the PM phase in Fig.~\ref{figure4} turns to be XY-FM phase~\cite{PhysRevLett.106.105302}, which fulfills $\langle S_i^z\rangle=0$ but presents a non-zero transversal magnetization.

From Fig.~\ref{figure4}, we read out that the boundary is $q=2z(J_1-J_2)$ for ferromagnetic case, more detailed calculation including the fluctuations of occupation probabilities gives the boundary as
\begin{eqnarray}
q_c=2z(D_0-D_1)(J_1-J_2).
\end{eqnarray}

Then $D_0$ and $D_1$ are determined by the self-consistent equations
\begin{eqnarray}
D_1&=&D_{01}\frac{1}{N_s}\sum_{\mathbf{k}}\left[F_{+}(\mathbf{k})+F_{-}(\mathbf{k})\right],\\
D_0&=&1-2D_1,
\end{eqnarray}
with
\begin{eqnarray}
F_{\pm}(\mathbf{k})=\frac{\pm\omega_{\mathbf{k},q}+q+zJ_2D_{01}+J_2D_{01}\zeta(\mathbf{k})}{2\omega_{\mathbf{k},q}}f(\pm\omega_{\mathbf{k},q}),\nonumber\\
\end{eqnarray}
where $\omega_{\mathbf{k},q}$ is defined as
\begin{eqnarray}
&&\omega_{\mathbf{k},q}\nonumber\\
&=&\sqrt{[q+D_{01}(J_1\zeta(\mathbf{k})+zJ_2)]^2-(J_2-J_1)^2D_{01}^2\zeta(\mathbf{k})^2}.\nonumber\\
\end{eqnarray}

\begin{figure}[htb]
\includegraphics[width=\columnwidth]{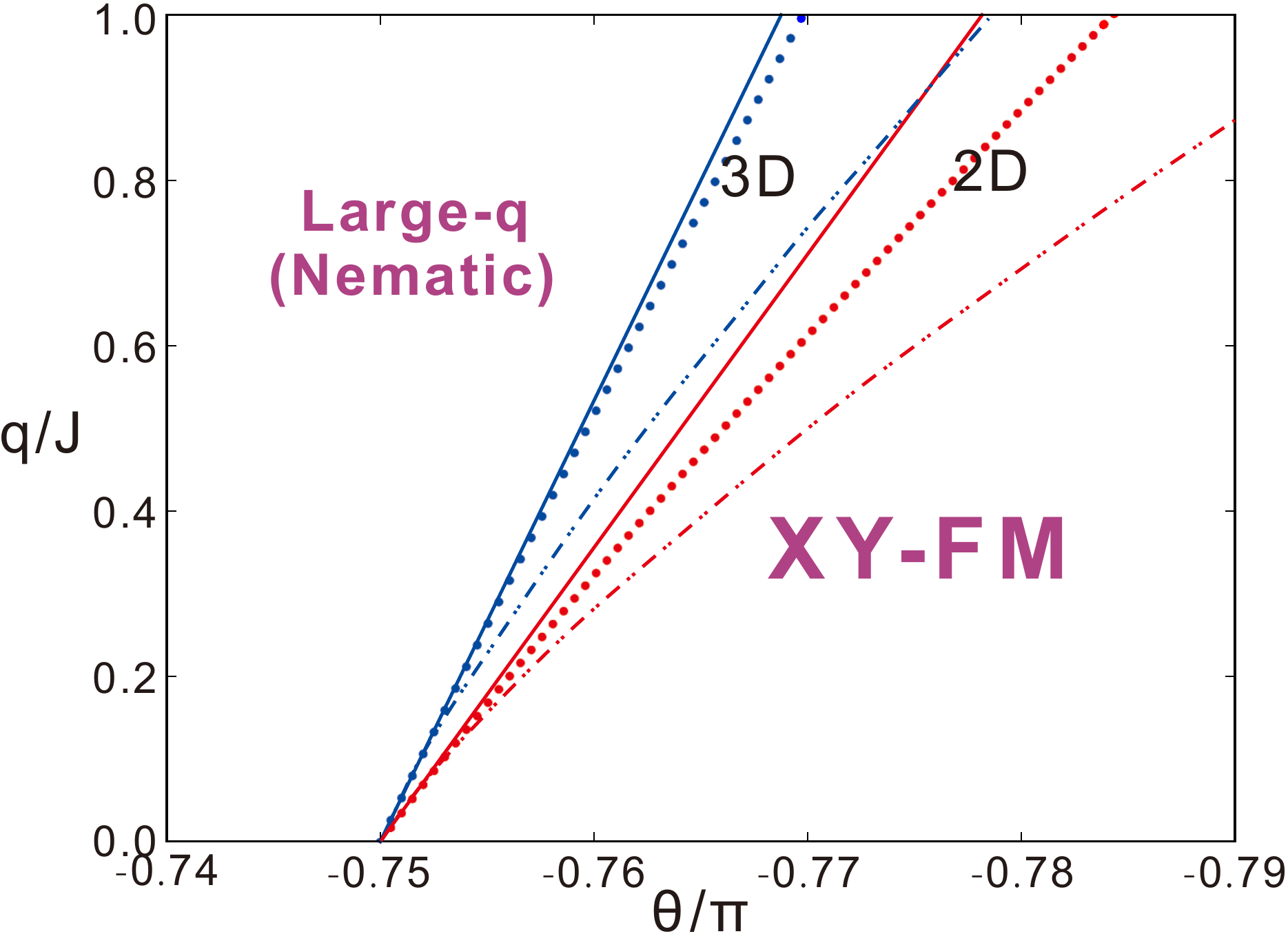}
\caption{Phase separation in the case $\lambda=0$. The solid curves represent analytical solutions without fluctuations, the dot-dashed curves represent SC-SBO results and the dot-segment-dashed curves represent the results calculated in Ref.~\cite{PhysRevLett.106.105302}. Here we have defined $J_1=-J\cos{\theta}$, $J_2=-J\sin{\theta}$ with $J=\sqrt{J_1^2+J_2^2}$.}
\label{figure5}
\end{figure}

In Fig.~\ref{figure5}, we give the analytical and SC-SBO phase separation diagrams as well as the result estimated in Ref.~\cite{PhysRevLett.106.105302}. We find that when $J_1\approx J_2$ ($\theta\approx-0.75\pi$), the results of SBO method and Ref.~\cite{PhysRevLett.106.105302} are very close.

\section{insulating state with two atoms per site}

\label{section_Mott2}

For insulating state with two atoms per site, the spin value in each site can be $S=0$ or $S=2$. For antiferromagnetic case ($U_2>0$), in the limit $t=0$, the on-site energy is minimized when $S=0$, corresponding to the singlet state. But when $t$ is of order $\sqrt{U_0U_2}$, we need to consider the excitation to $S=2$, of which the exchange energy is of order $t^2/U_0$.

Similar to insulating state with one atom case, we derive the effective Hamiltonian with perturbation theory ($t\ll U_0$), which can be expressed as
\begin{eqnarray}
\hat{H}_{\text{eff},2}&=&\sum_{\langle ij\rangle}\sum_{\alpha\alpha^\prime\beta\beta^\prime}H^{ij}_{\alpha\alpha^\prime,\beta\beta^\prime}\hat{L}^i_{\alpha\alpha^\prime}\hat{L}^j_{\beta\beta^\prime},
\label{Hamiltonian_eff_2}
\end{eqnarray}
where $|\alpha\rangle$-$|\beta^\prime\rangle$ belong to the sets $\{S=0\}$ and $\{S=2,S_z=-2,...,2\}$. $H_{\alpha\alpha^\prime,\beta\beta^\prime}$ can be derived with perturbation theory~\cite{PhysRevA.68.063602}. For simplicity, in the following consideration in this section, we label $|s\rangle=|0,0;2\rangle$ and $|m\rangle=|2,m;2\rangle$. In appexdix.D, we give detailed expressions for $H_{\alpha\alpha^\prime,\beta\beta^\prime}$.

When $t=0$, the system is initially in the singlet state $|s\rangle$, then we need to consider the $|s\rangle$ to $|m\rangle$ ($m=\pm2,\pm1,0$) excitations when $t$ is nonzero. We find the excitation spectrums are the same for different $m$ due to global SU(2) symmetry, so here we consider $|s\rangle$ to $|0\rangle$ excitation. Neglecting quantum fluctuations, the equations connecting to Green's functions $G_{s0,0s}(\mathbf{k},\omega)$ can be written as
\begin{eqnarray}
\mathbf{M}_{s\rightarrow 0}\begin{pmatrix}G_{s0,0s}(\mathbf{k},\omega) \\ G_{0s,0s}(\mathbf{k},\omega)\end{pmatrix}=\begin{pmatrix} D_{s0}\\ 0\end{pmatrix},
\end{eqnarray}
with
\begin{eqnarray}
\mathbf{M}_{s\rightarrow 0}&=&\begin{pmatrix}\zeta(\mathbf{k})H_{0s,s0} & \zeta(\mathbf{k})H_{0s,0s}\\
-\zeta(\mathbf{k})H_{s0,s0} & -\zeta(\mathbf{k})H_{s0,0s}\end{pmatrix}\nonumber\\
&&+\begin{pmatrix}\omega+E_s-E_0 & 0\\
0 & \omega+E_{0}-E_s\end{pmatrix},
\end{eqnarray}
where $E_s=zH_{ss,ss}$ and $E_0=zH_{00,ss}$. Then the excitation spectrum is given by
\begin{eqnarray}
\omega(\mathbf{k})&=&\sqrt{9U_2^2+16\frac{U_2t^2}{U_0}\zeta(\mathbf{k})},
\end{eqnarray}
which is the same as that derived in~\cite{PhysRevA.68.063602}. And by setting excitation gap to zero, we obtain the critical hopping energy as $t_c=\frac{3}{4}\sqrt{\frac{U_0U_2}{z}}$.

If we further include linear Zeeman effect (LZE) when external magnetic field is applied to the system~\cite{PhysRevLett.93.120405}, the effective Hamiltonian is updated to be
\begin{eqnarray}
\hat{H}_{\text{eff},2}=\sum_{i\alpha}V^{i}_{\alpha}\hat{L}^i_{\alpha\alpha}+\sum_{\langle ij\rangle}\sum_{\alpha\alpha^\prime\beta\beta^\prime}H^{ij}_{\alpha\alpha^\prime,\beta\beta^\prime}\hat{L}^i_{\alpha\alpha^\prime}\hat{L}^j_{\beta\beta^\prime},
\end{eqnarray}
with $V^i_{\alpha}=-\lambda S^z_{i\alpha}$. The result is symmetric for $\pm\lambda$, thus we restrict $\lambda>0$ in the following consideration.

In the limit of zero hopping energy $t=0$, the ground states can be $|s\rangle$ (singlet phase) when $\lambda<3/2U_2$ or $|2\rangle$ (ferromagnetic phase) when $\lambda>3/2U_2$. Start from singlet phase, the system can be excited to nematic phase when the hopping energy is larger than a critical value.  However, when $\lambda$ is large enough, the system is stable in FM phase without transition to other phases.

If $\lambda<3/2U_2$, and start from singlet phase, the system excite to state $|2\rangle$ first. Neglecting quantum fluctuations and assume $D_{\mu}=0$ for $\mu\neq s$, the critical hopping energy can be solved (see appendix.D) as
\begin{eqnarray}
t_c=\sqrt{\frac{U_0}{16z}(9U_2^2-4\lambda^2)}.
\end{eqnarray}

Similarly, start from FM phase, the system will be excited to state $|s\rangle$ and $|0\rangle$ first, and the boundary of the excitation without fluctuations is given by
\begin{eqnarray}
\frac{zt^2}{U_0}=\frac{3\lambda U_2-2\lambda^2}{8(\lambda-U_2)}.
\end{eqnarray}

\begin{figure}[htb]
\includegraphics[width=\columnwidth]{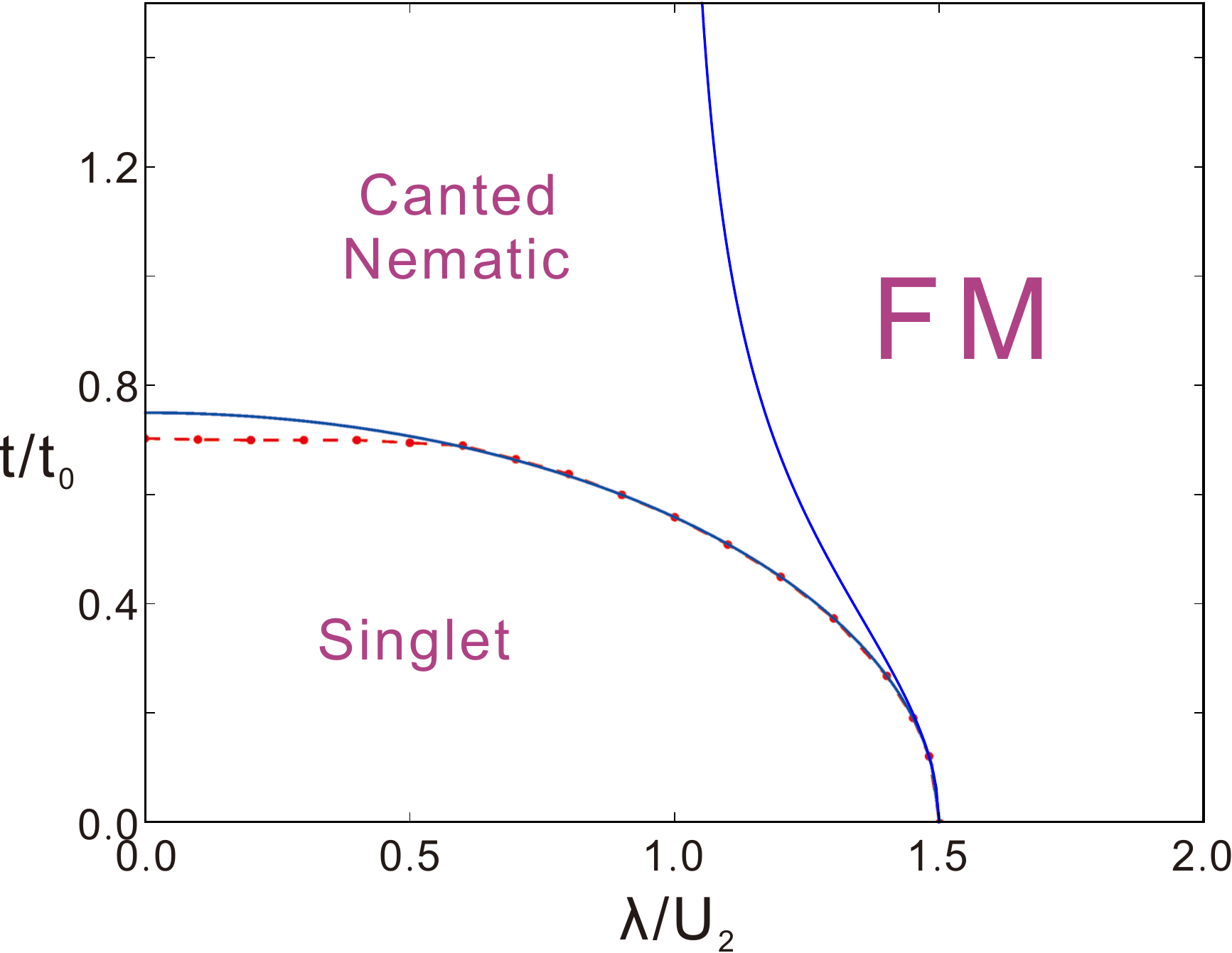}
\caption{Magnetic phase diagram in the Mott lobe with two atoms per site. The solid curve represent analytical result without quantum fluctuations, and the dashed curve represents the self-consistent result for singlet to nematic transition including quantum fluctuation. We have defined $t_0=\sqrt{U_0U_2/z}$ in the figure.}
\label{figure6}
\end{figure}

In Fig.~\ref{figure6}, we give the magnetic phase diagram for insulating phase with two atoms per site with LZE. The phase diagram can be interpreted as that when the external magnetic field is small, the system can be singlet or nematic phase, and there is also a small ferromagnetic component along the axis of the magnetic field, and as magnetic field goes up, the magnetization also increases and finally saturates in the FM regime. The nematic phase in the phase diagram is referred as canted nematic as the magnetization changes continuously in the phase~\cite{PhysRevLett.93.120405}. We also note that for $\lambda=0$ in Fig.~(\ref{figure6}), the self-consistent result give the critical hopping energy as $t_c\approx 0.7t_0$, which is very close to the global transition point given by mean-field~\cite{PhysRevA.68.063602} and Monte-Carlo simulation~\cite{PhysRevB.88.104509}.

\section{Conclusion}

\label{section_conlusion}

In this paper, we solve spin-1 bosonic system trapped in optical lattice with extended standard basis operator method. Both the MI-SF transition and magnetic phase diagrams in the insulating states are studied.

For ferromagnetic interacting case ($U_2<0$), we analytically calculate that both superfluid and Mott-insulator phases in the phase diagram are ferromagnetic. While for antiferromagnetic case ($U_2>0$), the superfluid phase is in polar state and the global magnetization in the whole phase diagram is zero.

Then we take quantum and thermal fluctuations of occupation probabilities in the system into account with self-consistent SBO approach and derive self-consistent SBO phase diagrams, which are in relevant agreement with Monte-Carlo simulation results at zero temperature. At finite temperature, we also compare the self-consistent SBO phase diagrams with mean-field phase diagrams and conclude that the finite-temperature-effect has been underestimated in mean-field approximation.

For $n=1$ Mott lobe with linear and quadratic Zeeman effect, we describe the system with an effective bilinear-biquadratic spin Hamiltonian. By considering the spin-excitations in this system with SBO method, we can decide the spin phase separations in the magnetic phase diagrams. For antiferromagnetic case, only nematic and ferromagnetic phases are expected, while for ferromagnetic case, another partial-nematic phase exists between nematic and ferromagnetic phases.

For $n=2$ Mott lobe with linear Zeeman effect, we apply similar approach and derive the magnetic phase diagram consisting of singlet, ferromagnetic and canted-nematic phases.

Besides the spin-1 system, the SBO method used in this paper can be extended to solve arbitrary spin-$S$ systems with similar procedures. However, there are still some cases, such as system in one dimensional lattice with broken translational symmetry in a regime, where the SBO method fails to explain the phenomenon.

\section{appendix}

\subsection{motion equation of Green's function}
\label{Append_A}
\setcounter{equation}{0}
\renewcommand{\theequation}{A.\arabic{equation}}

In this section, we will show how to derive the motions equation in SBO method. Then we will also calculate the Green's function matrix involved in Sec.~\ref{section_MI-SF} with the motion equations.

For an arbitrary Hamiltonian with the form
\begin{eqnarray}
\hat{H}=\sum_{i\mu\mu^\prime} V_{\mu\mu^\prime}^i \hat{L}^i_{\mu\mu^\prime} + \sum_{\langle ij\rangle}\sum_{\mu\mu^\prime\nu\nu^\prime}T^{ij}_{\mu\mu^\prime\nu\nu^\prime}\hat{L}^i_{\mu\mu^\prime}\hat{L}^j_{\nu\nu^\prime},
\label{Hamiltonian_form}
\end{eqnarray}
we can give the motion equation for Green's function $G^{ij}_{\mu\mu^\prime,\nu\nu^\prime}(t-t^\prime)=-i\Theta(t-t^\prime)\langle [\hat{L}^i_{\mu\mu^\prime}(t),\hat{L}^j_{\nu\nu^\prime}(t^\prime)]\rangle$.

Differentiating $G^{ij}_{\mu\mu^\prime,\nu\nu^\prime}(t-t^\prime)$ with respect to $t$ and Fourier transforming into $\omega$, we obtain the motion equation
\begin{eqnarray}
\omega G^{ij}_{\mu\mu^\prime,\nu\nu^\prime}(\omega)=\langle [\hat{L}^i_{\mu\mu^\prime},\hat{L}^j_{\nu\nu^\prime}]\rangle+ \langle\langle[\hat{L}^i_{\mu\mu^\prime},H]|\hat{L}^j_{\nu\nu^\prime}\rangle\rangle_{\omega}.
\label{Green_motion1}\nonumber\\
\end{eqnarray}

The last term in Eq.~(\ref{Green_motion1}) involves higher-order Green's function. For further calculation, we need to decouple these terms with two-operator Green's functions, so we take random-phase approximation (RPA)~\cite{PhysRevB.5.1106}, with which the three-operator Green's function can be expanded as
\begin{eqnarray}
&&\langle\langle \hat{L}^n_{\mu\mu^\prime}\hat{L}^l_{\xi\xi^\prime}|\hat{L}^m_{\nu\nu^\prime}\rangle\rangle\nonumber\\
&=&\delta_{\mu\mu^\prime}\langle \hat{L}^n_{\mu\mu}\rangle G^{lm}_{\xi\xi^\prime,\nu\nu^\prime}(\omega)+\delta_{\xi\xi^\prime} \langle \hat{L}^l_{\xi\xi}\rangle G^{nm}_{\mu\mu^\prime,\nu\nu^\prime}(\omega)\nonumber\\
&&+\delta^{nl}[\delta_{\xi\mu^\prime}G^{nm}_{\mu\xi^\prime,\nu\nu^\prime}(\omega)- \delta_{\mu\xi^\prime}G^{nm}_{\xi\mu^\prime,\nu\nu^\prime}(\omega)],
\end{eqnarray}
then combined with the equation
\begin{eqnarray}
[\hat{L}^m_{\mu\mu^\prime},\hat{L}^n_{\nu\nu^\prime}]=\delta^{mn}(\delta_{\nu\mu^\prime}\hat{L}^n_{\mu\nu^\prime}-\delta_{\mu\nu^\prime}\hat{L}^n_{\nu\mu^\prime}),
\end{eqnarray}
The motion equation in Eq.~(\ref{Green_motion1}) can be rewritten as
\begin{widetext}
\begin{eqnarray}
\omega G^{ij}_{\mu\mu^\prime,\nu\nu^\prime}(\omega)&=&\delta^{ij}\delta_{\mu\nu^\prime}\delta_{\nu\mu^\prime} D_{\mu\mu^\prime}^i+\sum_l\sum_{\xi\xi^\prime} D_{\mu\mu^\prime}^i T^{il}_{\mu^\prime\mu\xi\xi^\prime}G^{lj}_{\xi\xi^\prime,\nu\nu^\prime}(\omega)+\sum_{\alpha}\Big[\Big(V^i_{\mu^\prime\alpha}+\sum_{l}\sum_{\beta}D^{l}_{\beta}T^{il}_{\mu^\prime\alpha\beta\beta}\Big)G^{ij}_{\mu\alpha,\nu\nu^\prime}(\omega)\nonumber\\
&&-\Big(V^i_{\alpha\mu}+\sum_{l}\sum_{\beta}D^{l}_{\beta}T^{il}_{\alpha\mu\beta\beta}\Big)G^{ij}_{\alpha\mu^\prime,\nu\nu^\prime}(\omega)\Big].
\label{Green_motion2}
\end{eqnarray}
\end{widetext}

If we take $V^i_{\mu\mu^\prime}=\delta_{\mu\mu^\prime}V^i_{\mu}$ and $T^{ij}_{\mu\mu^\prime\nu\nu}=0$ for $\mu\neq\mu^\prime$ in Eq.~(\ref{Green_motion2}) (MI-SF transition and insulating states with one atom per site), the equation can be simplified by setting $E^i_{\mu}=V^i_\mu+z\sum_{\nu}D_{\nu}^iT_{\mu\mu\nu\nu}$.

Then to calculate the Green's function matrix $\mathbf{G}(\mathbf{k},\omega)$ introduced in Sec.~\ref{section_MI-SF}, we combine Eq.~(\ref{Green_motion}) and Eq.~(\ref{Green_alpha_beta}) and have
\begin{eqnarray}
&&\mathbf{G}(\mathbf{k},\omega)-\epsilon(\mathbf{k}) \mathbf{N}_{11}(\omega)\mathbf{G}(\mathbf{k},\omega)\nonumber\\
&=&\mathbf{N}_{11}(\omega)+\epsilon(\mathbf{k}) \mathbf{N}_{12}(\omega)\mathbf{G}^\prime(\mathbf{k},\omega),\\
&&\mathbf{G}^\prime(\mathbf{k},\omega)-\epsilon(\mathbf{k})\mathbf{N}_{22}(\omega)\mathbf{G}^\prime(\mathbf{k},\omega)\nonumber\\
&=&\mathbf{N}_{21}(\omega)+\epsilon(\mathbf{k})\mathbf{N}_{21}(\omega)\mathbf{G}(\mathbf{k},\omega),
\end{eqnarray}
where the matrix $\mathbf{G}^\prime(\mathbf{k},\omega)$ is defined as
\begin{eqnarray}
[\mathbf{G}^\prime]_{\alpha\beta}(\mathbf{k},\omega)=\sum_{\mu\mu^\prime\nu\nu^\prime} d^\alpha_{\mu\mu^\prime}d^\beta_{\nu\nu^\prime}G_{\mu\mu^\prime,\nu\nu^\prime}(\mathbf{k},\omega).
\end{eqnarray}

So finally we can solve $\mathbf{G}(\mathbf{k},\omega)$ as Eq.~(\ref{Green_matrix}).

\subsection{determination of the symmetry properties in the MI-SF phase transition}
\setcounter{equation}{0}
\renewcommand{\theequation}{B.\arabic{equation}}

In this section, we will give detailed analytical calculations to determine the symmetry properties in the MI-SF phase transition process. Obviously, we need to consider ferromagnetic and antiferromagnetic cases separately.

\subsubsection{ferromagnetic case ($U_2<0$)}

We find the coefficients satisfy the relationships
\begin{eqnarray}
&&\sum_{\alpha}\sum_s|c_s|^2\left[(A^{(\alpha)s,s+\alpha}_{n,n-1})^2+(A^{(\alpha)s,s+\alpha}_{n,n+1})^2\right]\nonumber\\
&=&\sum_{\alpha}\langle\psi_0|\hat{a}_{\alpha}\hat{a}_{\alpha}^\dagger|\psi_0\rangle\nonumber\\
&=&n+3,
\end{eqnarray}
\begin{eqnarray}
&&\sum_{\alpha}\sum_s|c_s|^2\left[(B^{(\alpha)s,s-\alpha}_{n,n-1})^2\right]\nonumber\\
&=&\sum_{\alpha}\langle\psi_0|\hat{a}_{\alpha}^\dagger\hat{a}_{\alpha}|\psi_0\rangle\nonumber\\
&=&n,
\end{eqnarray}
and
\begin{eqnarray}
&&\sum_{n^\prime=n\pm1}\sum_s c_{s-\alpha_1}c_{s-\alpha_2}^*A^{(\alpha_1)s-\alpha_1,s}_{n,n^\prime}A^{(\alpha_2)s-\alpha_2,s}_{n,n^\prime}\nonumber\\
&=&\langle\psi_0|\hat{a}_{\alpha_2}\hat{a}_{\alpha_1}^\dagger|\psi_0\rangle=\langle\psi_0|\hat{a}_{\alpha_1}^\dagger\hat{a}_{\alpha_2}|\psi_0\rangle~~\text{for $\alpha_1\neq\alpha_2$}\nonumber\\
&=&\sum_s c_{s+\alpha_2}c_{s+\alpha_1}^*B^{(\alpha_1)s+\alpha_1,s}_{n,n-1}B^{(\alpha_2)s+\alpha_2,s}_{n,n-1}.
\end{eqnarray}
Then the matrix $\mathbf{N}_{11}$ can be expanded as
\begin{eqnarray}
\mathbf{N}_{11}&=&(\Omega_2+\Omega_3)\mathbf{M}_{1}^\prime+\sum_s(\Omega_1-\Omega_2)\mathbf{M}_{2s}^\prime\nonumber\\
&&+\Omega_2\mathbf{I}_{3\times3},
\end{eqnarray}
where $\mathbf{M}_1^\prime$ and $\mathbf{M}_{2s}^\prime$ are defined as
\begin{eqnarray}
\mathbf{M}_1^\prime=\begin{pmatrix}\mathcal{M}_{0,0} &  \mathcal{M}_{0,1} & \mathcal{M}_{0,1}\\ \mathcal{M}_{1,0} & \mathcal{M}_{1,1} & \mathcal{M}_{1,-1}\\
\mathcal{M}_{-1,0} & \mathcal{M}_{-1,1} & \mathcal{M}_{-1,-1}
\end{pmatrix},
\end{eqnarray}
with $\mathcal{M}_{\alpha_1,\alpha_2}=\sum_s c_{s+\alpha_1}c^*_{s+\alpha_2}B^{(\alpha_1)s+\alpha_1,s}_{n,n-1}B^{(\alpha_2)s+\alpha_2,s}_{n,n-1}$, and
\begin{eqnarray}
\mathbf{M}_{2s}^\prime=\begin{pmatrix}\mathcal{M}^s_{0,0} & \mathcal{M}^s_{0,1} & \mathcal{M}^s_{0,-1}\\ \mathcal{M}^s_{1,0} & \mathcal{M}^s_{1,1} & \mathcal{M}^s_{1,-1}\\
\mathcal{M}^s_{-1,0} & \mathcal{M}^s_{-1,1} & \mathcal{M}^s_{-1,-1}
\end{pmatrix},
\end{eqnarray}
with $\mathcal{M}^s_{\alpha_1,\alpha_2}=c_{s-\alpha_1}c^*_{s-\alpha_2}A^{(\alpha_1)s-\alpha_1,s}_{n,n-1}A^{(\alpha_2)s-\alpha_2,s}_{n,n-1}$.

So $\lambda_m\mathbf{I}_{3\times 3}-\mathbf{N}_{11}$ can be expanded as the summation of a series of hermite matrices
\begin{eqnarray}
&&\lambda_m\mathbf{I}_{3\times 3}-\mathbf{N}_{11}\nonumber\\
&=&(\Omega_2+\Omega_3)(n\mathbf{I}_{3\times 3}-\mathbf{M}_1^\prime)+\sum_s(\Omega_2-\Omega_1)\mathbf{M}_{2s}^\prime.
\label{matrix_expand}\nonumber\\
\end{eqnarray}

To prove the positive-semidefinite property of a matrix, we need to calculate the $3$ upper left determinants. Before the calculation, we give some inequality relations.
\begin{widetext}
\begin{eqnarray}
&&\mathcal{M}_{\alpha_1,\alpha_1}\mathcal{M}_{\alpha_2,\alpha_2}-\mathcal{M}_{\alpha_1,\alpha_2}\mathcal{M}_{\alpha_2,\alpha_1}\nonumber\\
&=&\sum_{s_1}|P_B^{(\alpha_1)}(s_1)|^2\sum_{s_2}|P_B^{(\alpha_2)}(s_2)|^2-\sum_{s_1,s_2}P_B^{(\alpha_1)}(s_1)[P_B^{(\alpha_2)}(s_1)]^*[P_B^{(\alpha_1)}(s_2)]^*P_B^{(\alpha_2)}(s_2)\nonumber\\
&=&\frac{1}{2}\sum_{s_1,s_2}\left|P_B^{(\alpha_1)}(s_1)P_B^{(\alpha_2)}(s_2)-P_B^{(\alpha_2)}(s_1)P_B^{(\alpha_1)}(s_2)\right|^2\nonumber\\
&\geq&0,
\end{eqnarray}
\begin{eqnarray}
&&\mathcal{M}_{1,1}\mathcal{M}_{0,-1}\mathcal{M}_{-1,0}+\mathcal{M}_{-1,-1}\mathcal{M}_{0,1}\mathcal{M}_{1,0}-\mathcal{M}_{1,-1}\mathcal{M}_{0,1} \mathcal{M}_{-1,0}-\mathcal{M}_{-1,1} \mathcal{M}_{0,-1}\mathcal{M}_{1,0}\nonumber\\
&=&\sum_{s}\left[|P^{(1)}_B(s)|^2|\mathcal{M}_{0,-1}|^2+|P^{(-1)}_B(s)|^2|\mathcal{M}_{0,1}|^2-P^{(1)}_B(s)[P^{(-1)}_B(s)]^*\mathcal{M}_{0,1}\mathcal{M}_{0,-1}^* -[P^{(1)}_B(s)]^*P^{(-1)}_B(s)\mathcal{M}_{0,-1}\mathcal{M}_{0,1}^*\right]\nonumber\\
&=&\sum_s\left|P^{(1)}_B(s)\mathcal{M}_{0,-1}^*-P^{(-1)}_B(s)\mathcal{M}_{0,1}^*\right|^2\nonumber\\
&\geq&0,
\end{eqnarray}
\end{widetext}
where $P_B^{(\alpha)}(s)$ is defined in Eq.~(\ref{P_B_definition}).

For matrix $n\mathbf{I}_{3\times 3}-\mathbf{M}_1^\prime$, we have
\begin{widetext}
\begin{eqnarray}
\text{UL}_1=n-\mathcal{M}_{0,0}=n-\langle\psi_0|\hat{a}_0^\dagger\hat{a}_0|\psi_0\rangle\geq 0,
\end{eqnarray}
\begin{eqnarray}
\text{UL}_2&=&(n-\mathcal{M}_{0,0})(n-\mathcal{M}_{1,1})-\mathcal{M}_{0,1}\mathcal{M}_{1,0}\nonumber\\
&\geq&n(n-\mathcal{M}_{0,0}-\mathcal{M}_{1,1})\nonumber\\
&\geq&0,
\end{eqnarray}
\begin{eqnarray}
\text{UL}_3&=&(n-\mathcal{M}_{0,0})[(n-\mathcal{M}_{1,1})(n-\mathcal{M}_{-1,-1})-\mathcal{M}_{1,-1}\mathcal{M}_{-1,1}]\nonumber\\
&&+\mathcal{M}_{0,1}[-\mathcal{M}_{1,0}(n-\mathcal{M}_{-1,-1})-\mathcal{M}_{1,-1}\mathcal{M}_{-1,0}]\nonumber\\
&&-\mathcal{M}_{0,-1}[\mathcal{M}_{1,0}\mathcal{M}_{-1,1}+\mathcal{M}_{-1,0}(n-\mathcal{M}_{1,1})]\nonumber\\
&\geq&n(n-\mathcal{M}_{0,0})(n-\mathcal{M}_{1,1}-\mathcal{M}_{-1,-1})-n\mathcal{M}_{0,0}\mathcal{M}_{1,1}-n\mathcal{M}_{0,0}\mathcal{M}_{-1,-1}\nonumber\\
&=&0.
\label{UL_3}
\end{eqnarray}
\end{widetext}

We have applied the relation $\mathcal{M}_{0,0}+\mathcal{M}_{1,1}+\mathcal{M}_{-1,-1}=n$ in the calculation of $\text{UL}_l$ ($l=1,2,3$). So matrix $n\mathbf{I}_{3\times 3}-\mathbf{M}_1^\prime$ is positive-semidefinite.

Also for matrix $\mathbf{M}_{2s}^\prime$, we calculate $\text{UL}^s_{1}\geq0$ and $\text{UL}^s_{2}=\text{UL}^s_{3}=0$. So combined with the conditions $\Omega_2+\Omega_3>0$ and $\Omega_2-\Omega_1>0$, we conclude from Eq.~(\ref{matrix_expand}) that $\lambda_m\mathbf{I}_{3\times 3}-\mathbf{N}_{11}$ is positive-semidefinite, which indicates that the eigenvalues of $\mathbf{N}_{11}$ satisfy the condition $\lambda\leq\lambda_m$.

Then we need to consider the conditions for $\lambda=\lambda_m$. Based on the Minkowski determinant theorem~\cite{Marcus1964A}, $\text{UL}_3$ in Eq.~(\ref{UL_3}) has to equal to zero, which indicates
\begin{eqnarray}
\frac{P^{(\alpha_1)}_B(s)}{P^{(\alpha_2)}_B(s)}=R_{\alpha_1,\alpha_2},
\end{eqnarray}
where $R_{\alpha_1\alpha_2}$ is an $s$-independent number. And we assume that $P_B^{(\alpha_2)}(s)\neq0$.

Moreover, we note that the eigenvectors of matrices $n\mathbf{I}_{3\times 3}-\mathbf{M}_1^\prime$ and $\mathbf{M}_{2s}^\prime$ are the same, which would induce
\begin{eqnarray}
1+R_{1,0}r_{1,0}+R_{-1,0}r_{-1,0}=0,
\label{FM_condition1}
\end{eqnarray}
where we have also defined
\begin{eqnarray}
\frac{P^{(\alpha_1)}_A(s)}{P^{(\alpha_2)}_A(s)}=r_{\alpha_1,\alpha_2}.
\end{eqnarray}

Also we note that $R_{1,0}=-r_{-1,0}$ because
\begin{eqnarray}
&&\frac{A^{(-1)s+1,s}_{n,n-1}}{A^{(0)s,s}_{n,n-1}}=\frac{\langle n-1,s;n-1|\hat{\Theta}\hat{a}_{-1}^\dagger|n,s+1;n\rangle}{\langle n-1,s;n-1|\hat{\Theta}\hat{a}_{0}^\dagger|n,s;n\rangle}\nonumber\\
&=&\frac{\langle n-1,s;n-1|\hat{a}_{-1}^\dagger\hat{\Theta}-2\hat{a}_1|n,s+1;n\rangle}{\langle n-1,s;n-1|\hat{a}_{0}^\dagger\hat{\Theta}+2\hat{a}_0|n,s;n\rangle}\nonumber\\
&=&-\frac{\langle n-1,s;n-1|\hat{a}_1|n,s+1;n\rangle}{\langle n-1,s;n-1|\hat{a}_0|n,s;n\rangle}\nonumber\\
&=&-\frac{B^{(1)s+1,s}_{n,n-1}}{B^{(0)s,s}_{n,n-1}},
\end{eqnarray}
where operator $\hat{\Theta}=\hat{a}_0^2-2\hat{a}_1\hat{a}_{-1}$ describes the destruction of a spin singlet pair.

Similarly, we have $R_{-1,0}=-r_{1,0}$. So Eq.~(\ref{FM_condition1}) is equivalent to
\begin{eqnarray}
R_{1,0}R_{-1,0}=\frac{1}{2}.
\label{FM_condition2}
\end{eqnarray}

Then we can calculate $\langle \vec{S}\rangle^2$ in the system
\begin{eqnarray}
&&\langle \vec{S}\rangle^2=\langle\hat{S}^+\rangle\langle\hat{S}^-\rangle+\langle\hat{S}_z\rangle^2\nonumber\\
&=&2|R_{1,0}+R_{-1,0}^*|^2\Big(\sum_s|P^{(0)}_B(s)|^2\Big)^2+\nonumber\\
&&(|R_{1,0}|^2-|R_{-1,0}|^2)^2\Big(\sum_s|P^{(0)}_B(s)|^2\Big)^2\nonumber\\
&=&n^2.
\end{eqnarray}

The corresponding eigenvector involved in the calculation, which can be expressed as $(1,R_{1,0},R_{-1,0})$, determines the symmetry properties of the superfluid phase. So we conclude from Eq.~(\ref{FM_condition2}) that the superfluid phase in the excitation is in ferromagnetic state.

\subsubsection{antiferromagnetic case ($U_2>0$)}

For odd filling number case, the matrix $\mathbf{N}_{11}(0)$ can be expressed as
\begin{eqnarray}
\mathbf{N}_{11}(0)=3(\Upsilon_2+\Upsilon_4)\mathbf{I}_{3\times3}+K_1\mathbf{M}_1+K_2\mathbf{M}_2,
\end{eqnarray}
where matrices $\mathbf{M}_1$ and $\mathbf{M}_2$ are
\begin{eqnarray}
\mathbf{M}_1&=&\begin{pmatrix}|c_0|^2 & c_0c_1^* & c_0c_{-1}^*\\c_0^*c_1 & |c_1|^2 & c_1c_{-1}^*\\ c_0^*c_{-1} & c_{-1}c_{1}^* & |c_{-1}|^2\end{pmatrix},
\end{eqnarray}
and
\begin{eqnarray}
\mathbf{M}_2&=&\begin{pmatrix}|c_0|^2 & -c_0^*c_{-1} & -c_0^*c_{1}\\-c_0c_{-1}^* & |c_{-1}|^2 & c_1c_{-1}^*\\ -c_0c_{1}^* & c_{-1}c_{1}^* & |c_{1}|^2\end{pmatrix}.
\end{eqnarray}

\subsection{analysis of the spin phase boundary in MI phase with one atom per site}
\setcounter{equation}{0}
\renewcommand{\theequation}{C.\arabic{equation}}

In this section, we give detailed analysis of the minimum value of Eq.~(\ref{det_1}), which would induce different phases in the MI phase with one atom per site.

For the excitation from $|1,0;1\rangle$ to $|1,1;1\rangle$, we start from Eq.~(\ref{det_1}) and define
\begin{eqnarray}
f(\eta)&=&(q+zJ_2+J_1\eta)^2-(J_2-J_1)^2\eta^2,
\end{eqnarray}
with $-z\leq \eta \leq z$. Then we calculate the minimum value of $f(\eta)$.

When $J_2\geq2J_1$, the minimum is obtained by comparing $f(z)$ and $f(-z)$, and we have
\begin{eqnarray}
f(\eta)\Big|_{\text{min}}=(q+zJ_2-zJ_1)^2-z^2(J_2-J_1)^2.
\end{eqnarray}

When $J_2<2J_1$, the case becomes more complicated, we need to first decide whether the global minimum point is within the range $[-z,z]$.

If $q\geq zJ_2-z\frac{J_2^2}{J_1}$, the minimum is on the point $\eta=-z$ with
\begin{eqnarray}
f(\eta)\Big|_{\text{min}}=(q+zJ_2-zJ_1)^2-z^2(J_2-J_1)^2,
\end{eqnarray}
and if $q<zJ_2-z\frac{J_2^2}{J_1}$, the minimum is on the global minimum point with
\begin{eqnarray}
f(\eta)\Big|_{\text{min}}=-\frac{(J_2-J_1)^2}{J_1^2-(J_2-J_1)^2}(q+zJ_2)^2<0.
\end{eqnarray}

\subsection{SBO method for MI phase with two atoms per site}
\setcounter{equation}{0}
\renewcommand{\theequation}{D.\arabic{equation}}

In this section, we give the self-consistent equations for the spin-excitation in the MI phase with two atoms per site. Then by solving the equations, we will derive self-consistent results of the phase transition process.

For MI phase with two atoms per site, $H_{\alpha\alpha^\prime,\beta\beta^\prime}$ in Eq.~(\ref{Hamiltonian_eff_2}) can be expressed as
\begin{eqnarray}
H_{ss,ss}&=&-\frac{20t^2}{3U_0},\\
H_{ms,sm^\prime}&=&-\frac{8t^2}{3U_0}\delta_{mm^\prime},\\
H_{mm^\prime, ss}&=&\Big(-\frac{20t^2}{3U_0}+\frac{3U_2}{z}\Big)\delta_{mm^\prime},\\
H_{ms,m^\prime s}&=&(-1)^{-m+1}\frac{8t^2}{3U_0}\delta_{m,-m^\prime},
\end{eqnarray}
\begin{eqnarray}
H_{sm, lm^\prime}&=&\frac{4\sqrt{7}t^2}{3U_0}C_{0,0;2,l}^{2,l}C_{2,m;2,m^\prime}^{2,l}\delta_{l,m+m^\prime},
\end{eqnarray}
\begin{eqnarray}
&&H_{ml, m^\prime l^\prime}\nonumber\\
&=&\delta_{m+m^\prime,l+l^\prime}\Big[\left(\frac{6U_2}{z}-\frac{16t^2}{3U_0}\right)C_{2,m;2,m^\prime}^{0,m+m^\prime}C_{2,l;2,l^\prime}^{0,l+l^\prime}\nonumber\\
&&+\left(\frac{6U_2}{z}-\frac{4t^2}{U_0}\right)C_{2,m;2,m^\prime}^{1,m+m^\prime}C_{2,l;2,l^\prime}^{1,l+l^\prime}\nonumber\\
&&+\left(\frac{6U_2}{z}-\frac{8t^2}{3U_0}\right)C_{2,m;2,m^\prime}^{2,m+m^\prime}C_{2,l;2,l^\prime}^{2,l+l^\prime}\nonumber\\
&&+\left(\frac{6U_2}{z}-\frac{4t^2}{U_0}\right)C_{2,m;2,m^\prime}^{3,m+m^\prime}C_{2,l;2,l^\prime}^{3,l+l^\prime}\nonumber\\
&&+\left(\frac{6U_2}{z}-\frac{12t^2}{U_0}\right)C_{2,m;2,m^\prime}^{4,m+m^\prime}C_{2,l;2,l^\prime}^{4,l+l^\prime}\Big].
\end{eqnarray}

In the above expressions, we have assumed that $|m\rangle,|l\rangle,|m^\prime\rangle,|l^\prime\rangle$ represent states that belong to set $\{S=2\}$. And $C_{S_1,m_1;S_2,m_2}^{S,m}$ are the known Clebsch-Gordon coefficients.

For singlet ($|s\rangle$) to ferromagnetic ($|2\rangle$) excitation, we give the equations connecting to $G_{s2,2s}(\mathbf{k},\omega)$
\begin{eqnarray}
\mathbf{M}_{s\rightarrow 2}\begin{pmatrix}G_{s2,2s}(\mathbf{k},\omega) \\ G_{-2s,2s}(\mathbf{k},\omega) \\ G_{02,2s}(\mathbf{k},\omega)\end{pmatrix}=\begin{pmatrix}D_{s2} \\ 0 \\ 0\end{pmatrix},
\end{eqnarray}
with
\begin{eqnarray}
&&\mathbf{M}_{s\rightarrow 2}=\omega \mathbf{I}_{3\times 3}+\begin{pmatrix}E_s-E_2 & 0 & \gamma\\ 0 & E_{-2}-E_s & 0 \\ \gamma & 0 & E_0-E_2\end{pmatrix}+\nonumber\\
&&\zeta(\mathbf{k})\begin{pmatrix}D_{s2}H_{s2,s2} & D_{s2}H_{2s,-2s} & D_{s2}H_{2s,02}\\
D_{-2s}H_{s-2,s2} & D_{-2s}H_{s-2,-2s} & D_{-2s}H_{s-2,02}\\
D_{02}H_{20,s2} & D_{02}H_{20,-2s} & D_{02}H_{20,02}\end{pmatrix},\nonumber\\
\end{eqnarray}
where $E_{\mu}=V_{\mu}+z\sum_{\nu}D_{\nu}H_{\mu\mu,\nu\nu}$ and $\gamma=z\sum_{\nu}D_{\nu}H_{0s,\nu\nu}$. We have eliminated $G_{-20,2s}(\mathbf{k},\omega)$ and $G_{-11,2s}(\mathbf{k},\omega)$ in the equations because $E_2<E_{m}$ ($m=-2,\pm1,0$) and we have $D_{\mu}=0$ for $\mu\neq s,2$ in the calculation. Then we can solve $D_2$ with integrations of the Green's functions.

\bibliography{mypaper}
\end{document}